# Loss functions arising from the index of agreement

Hristos Tyralis[1*], Georgia Papacharalampous[2]

[1]Support Command, Hellenic Air Force, Elefsina Air Base, 19 200, Elefsina, Greece (montchrister@gmail.com, hristos@itia.ntua.gr, https://orcid.org/0000-0002-8932-4997)

[2]Department of Land, Environment, Agriculture and Forestry, University of Padova, Viale dell'Università 16, 35020, Legnaro, Italy (papacharalampous.georgia@gmail.com, georgia.papacharalampous@unipd.it, https://orcid.org/0000-0001-5446-954X)

*Corresponding author

**Abstract**: We examine the theoretical properties of the index of agreement loss function $L_W$, the negatively oriented counterpart of Willmott's index of agreement, a common metric in environmental sciences and engineering. We prove that $L_W$ is bounded within [0,1], translation and scale invariant, and estimates the parameter $\mathbb{E}_F[\underline{y}] \pm \mathbb{V}_F^{1/2}[\underline{y}]$ when fitting a distribution. We propose $L_{\mathrm{NR}_2}$ as a theoretical improvement, which replaces the denominator of $L_W$ with the sum of Euclidean distances, better aligning with the underlying geometric intuition. This new loss function retains the appealing properties of $L_W$ but also admits closed-form solutions for linear model parameter estimation. We show that as the correlation between predictors and the dependent variable approaches 1, parameter estimates from squared error, $L_{\mathrm{NR}_2}$ and $L_W$ converge. This behavior is mirrored in hydrologic model calibration (a core task in water resources engineering), where performance becomes nearly identical across these loss functions. Finally, we suggest potential improvements for existing $L_p$-norm variants of the index of agreement.

**Keywords**: extremum estimation; index of agreement; loss function; model calibration; model evaluation; point forecasting

## 1. Introduction

Loss functions are fundamental tools in machine learning, statistical science and engineering, used for model estimation (also referred to as fitting, calibrating or training) and model ranking (evaluation) [1]. During fitting, model parameters are estimated by minimizing the average discrepancy between predictions and observations, quantified by the loss function. For ranking, a model's performance is evaluated by comparing the average loss of predictions against observations in a test set.

Domain-specific loss functions developed in environmental sciences and engineering ([2], [3], [4]) hold potential for broader application. One established metric is Willmotts's [5] index of agreement, defined as

$$d(\boldsymbol{z},\boldsymbol{y}) := 1 - \sum_{i=1}^{n}(z_i - y_i)^2 \,/\, \sum_{i=1}^{n}(|z_i - \mu(\boldsymbol{y})| + |\mu(\boldsymbol{y}) - y_i|)^2 \tag{1.1}$$

where $\boldsymbol{z} = (z_1, \dots, z_n)^{\mathrm{T}}$ represents model predictions, $\boldsymbol{y} = (y_1, \dots, y_n)^{\mathrm{T}}$ represents realizations and $\mu(\boldsymbol{y})$ is the sample mean of realizations (defined in eq. (A.19)). Since $d$ is positively oriented (higher values indicate better performance), we study its negatively oriented counterpart without loss of generality, which we name the index of agreement loss function $L_W$ (subscript $W$ for Willmott):

$$L_W(\boldsymbol{z},\boldsymbol{y}) := \sum_{i=1}^{n}(z_i - y_i)^2 \,/\, \sum_{i=1}^{n}(|z_i - \mu(\boldsymbol{y})| + |\mu(\boldsymbol{y}) - y_i|)^2 \tag{1.2}$$

Although the index of agreement has been extensively researched, most studies are empirical [6] and a theoretical analysis is lacking. Here we aim to provide a theoretical foundation for the index of agreement and to propose variants with improved theoretical properties, building upon its fundamental construction principles. Our contributions are as follows:

(i) We examine the theoretical properties of $L_W(\boldsymbol{z},\boldsymbol{y})$ as an extremum estimator [7], demonstrating that it is bounded in [0,1], and is translation and scale invariant. When fitting a sample of realizations $\boldsymbol{y}$, it estimates the parameter $\hat{\theta} = \mu(\boldsymbol{y}) \pm \sigma(\boldsymbol{y})$, where $\sigma(\boldsymbol{y})$ is the sample standard deviation defined in eq. (A.21). For linear models, no closed-form solution has been identified, which remains an open problem.

(ii) We propose an alternative definition for $L_W$, which we call $L_{\mathrm{NR}_2}$ (subscript NR for Norm Ratio, subscript 2 for $L_2$-norm), defined in Section 3.2. This function maintains consistency with the distance-based arguments underlying the index of agreement, but uses $L_2$-norm-based constructions. Thus, $L_{\mathrm{NR}_2}$ represents a theoretical improvement over the original index. It preserves the [0,1] bound and translation and scale invariance. When fitting a sample $\boldsymbol{y}$, it estimates $\hat{\theta} = \mu(\boldsymbol{y}) \pm \sigma(\boldsymbol{y})$, analogous to the original index. For linear models $\boldsymbol{z} = a\boldsymbol{x} + b\boldsymbol{1}$, we provide closed-form estimates for parameters $a$ and $b$.

(iii) Constructions based on $L_p$-norms (in Section 3.3) generalize the new index of agreement, enabling a theoretically consistent variant of Legates and McCabe Jr [8] and Krause et al. [9] index of agreement variants.

(iv) Although our theoretical developments focus on estimation, we empirically extend these insights to model evaluation. Such an investigation is necessary because a theory linking estimation and evaluation for non-additive extremum estimators (unlike additive $M$-estimators; see [10]) is currently lacking in statistical science. Empirical results based on predictions from simulated data and real-world case studies (the latter using hydrologic models) estimated with $L_W$ and $L_{\mathrm{NR}_2}$, will facilitate a reassessment of prior studies that employed the index of agreement.

The remainder of the manuscript is structured as follows. Section 2 reviews essential background on prediction evaluation, estimation theory and the index of agreement. Section 3 presents new material and the main theoretical results, focusing on the original index of agreement and our subsequent developments. Empirical demonstrations using both simulated and real-world data modeled by hydrologic models is included in Section 4. The implications of these theoretical contributions are discussed in Section 5, followed by concluding remarks in Section 6. Given the extensive vector representations and calculus required for the theoretical analysis and proofs, Appendix A introduces consistent notation and definitions for vector calculus. Common identification [1] and loss functions used throughout the manuscript, are detailed in Appendix B, while the probability distributions used are listed in Appendix C. All proofs are supplied in Appendix D. Appendix E details the statistical software to reproduce the analysis. Supplementary material to reproduce all empirical analyses and figures is available.

## 2. Theoretical background

Section 2.1 outlines the theoretical framework for prediction evaluation, which applies to additive formulations of a loss function for both assessing a single model and comparing multiple models. As the index of agreement is non-additive, its application to parameter estimation places it within the broader category of extremum estimators. The theoretical foundations for this general class of estimators, including the specific case of $M$-estimators (which align directly to the framework in Section 2.1), are presented in Section 2.2. Section 2.3 then describes the index of agreement and develops the formulations required for its subsequent analysis in Section 3.

## 2.1 Comparison and adequacy of predictions

### *2.1.1 Comparing predictions from multiple models*

Let $\boldsymbol{z} = (z_1, \dots, z_n)^{\mathrm{T}}$ represent a vector of model predictions and $\boldsymbol{y} = (y_1, \dots, y_n)^{\mathrm{T}}$ a vector of corresponding observed values. Model comparisons are performed using a loss function $L(z, y)$ which quantifies the discrepancy between a single prediction $z$ and its observation $y$ ([21], [1]). Standard loss functions include the absolute error (AE) loss $L_{\mathrm{AE}}$ (eq. (B.1)) and the squared error (SE) loss $L_{\mathrm{SE}}$ (eq. (B.2)). A model is preferred over another if it achieves a lower average loss $\bar{L}$ across all observations. This average is defined as [1]:

$$\bar{L}(\boldsymbol{z}, \boldsymbol{y}) = (1/n) \sum_{i=1}^{n} L(z_i, y_i) \tag{2.1}$$

Commonly used averages of these loss functions include the mean absolute error (MAE; eq. (B.3)), which corresponds to the AE loss and the mean squared error (MSE; eq. (B.4)), which corresponds to the SE loss.

### *2.1.2 Strictly consistent loss functions*

A key property a loss function can have is known as (strict) consistency. To explain this concept, we must first define the notion of a statistical functional. Let $\underline{y}$ be a random variable (hereinafter, random variables are underlined). A materialization of $\underline{y}$ is written as $y$. We write $\underline{y} \sim F$ to indicate that $\underline{y}$ follows a cumulative distribution function (CDF) $F$, defined as:

$$F(y) := P(\underline{y} \leq y) \tag{2.2}$$

A one-dimensional statistical functional (or simply a functional) $T$ is a mapping [1]

$$T: \mathcal{F} \to \mathcal{P}(D), F \mapsto T(F) \subseteq D \tag{2.3}$$

where $\mathcal{F}$ is a class of probability distributions. The functional maps each distribution $F \in \mathcal{F}$ to a subset $T(F)$ of the domain $D \subseteq \mathbb{R}$ (i.e., an element of the power set $\mathcal{P}(D)$). This subset $T(F)$ can be single-valued or set-valued. Common examples of statistical functionals include the median $Q^{1/2}(F)$, the $\tau$-quantile $Q^{\tau}(F)$ and the mean $\mathbb{E}_{F_{\underline{y}}}[\underline{y}]$.

A loss function $L$ is $\mathcal{F}$-consistent for the functional $T$ if the following inequality holds for all distributions in the class ([1], [11])

$$\mathbb{E}_F[L(t, \underline{y})] \leq \mathbb{E}_F[L(z, \underline{y})] \; \forall \; F \in \mathcal{F}, t \in T(F), z \in D \tag{2.4}$$

The loss function is strictly $\mathcal{F}$-consistent if it is $\mathcal{F}$-consistent and equality in eq. (2.4) implies that $z \in T(F)$.

The practical importance of strict consistency can be understood from the following example. Suppose a modeler is tasked with predicting the mean of the random variable $\underline{y}$. The SE loss $L_{\mathrm{SE}}$ is strictly consistent for the mean [1]. This means that if the modeler follows the directive, i.e. predicts the true mean, her/his predictions will, in expectation, minimize the loss when evaluated using $L_{\mathrm{SE}}$. Eq. (2.4) is a population-level property, therefore, its natural empirical counterpart for comparing predictions is the sample average loss defined in eq. (2.1). $L_{\mathrm{SE}}$ belongs to the broader family of Bregman loss functions, all of which are strictly consistent for the mean ([1], [12], [13], [14], [15], [16]).

*2.1.3 Skill scores*

When evaluating predictions from multiple competing models, skill scores provide a standardized framework for comparison. The general form of a skill score, as given by [1] is:

$$\bar{L}_{\mathrm{skill}}(\boldsymbol{z}, \boldsymbol{y}, \boldsymbol{z}_{\mathrm{ref}}) := (\bar{L}(\boldsymbol{z}, \boldsymbol{y}) - \bar{L}(\boldsymbol{z}_{\mathrm{ref}}, \boldsymbol{y}))/(\bar{L}(\boldsymbol{z}_{\mathrm{optimal}}, \boldsymbol{y}) - \bar{L}(\boldsymbol{z}_{\mathrm{ref}}, \boldsymbol{y})) \tag{2.5}$$

where $\boldsymbol{z}$ represents the predictions from the model of interest, $\boldsymbol{z}_{\mathrm{ref}}$ represents the predictions from the reference model and $\boldsymbol{z}_{\mathrm{optimal}}$ represents the theoretically optimal predictions that achieve the minimum possible loss. Skill scores of this form are positively oriented.

A common simplification arises when the loss function $L$ has a minimum of 0, as this implies $\bar{L}(\boldsymbol{z}_{\mathrm{optimal}}, \boldsymbol{y}) = 0$. The skill score then simplifies to:

$$\bar{L}_{\mathrm{skill}}(\boldsymbol{z}, \boldsymbol{y}, \boldsymbol{z}_{\mathrm{ref}}) := 1 - \bar{L}(\boldsymbol{z}, \boldsymbol{y})/\bar{L}(\boldsymbol{z}_{\mathrm{ref}}, \boldsymbol{y}) \tag{2.6}$$

This ratio-based form is intuitively interpretable as the relative improvement over the reference method.

A prominent example is the Nash-Sutcliffe efficiency (NSE) [17], a skill score based on the squared error loss relative to a climatological mean [18]. The NSE is constructed by taking $L_{\mathrm{SE}}$ as the evaluation loss and using the mean climatology as the reference prediction. Substituting these into eq. (2.6) yields the NSE, which is defined in eq. (B.5).

*2.1.4 Adequacy of predictions from a single model*

While loss functions are instrumental for the comparative ranking of models, a distinct framework is required to assess the absolute adequacy of a single set of predictions. This assessment is formalized through the concept of identification functions ([1], [19]).

An identification function $V: D \times I \to \mathbb{R}, I \subseteq \mathbb{R}$ serves as a tool for evaluating whether a prediction $z$ equals a specified functional $T(F)$. The function $V$ is termed a strict $\mathcal{F}$-identification function for the functional $T(F)$ if it satisfies the following condition

$$\mathbb{E}_F[V(z, \underline{y})] = 0 \Leftrightarrow z \in T(F)\ \forall\ F \in \mathcal{F}, t \in T(F), z \in D \tag{2.7}$$

If a strict identification function exists for the functional $T$, then $T$ is called identifiable. The practical significance of identifiability lies in its ability to enable an absolute check of predictive performance. For instance, if a model claims to predict the median, the average of its identification function values over a test set can be used to statistically verify this claim, a concept aligned with calibration in the statistical literature ([18], [20]).

It is crucial to distinguish the roles of loss and identification functions. The former are used for comparative model ranking, while the latter are used for the absolute assessment of predictions for a specific functional [21].

A notable connection between these concepts is established by Osband's principle [22]. This principle states that a strictly $\mathcal{F}$-consistent loss function $L$ and a strict $\mathcal{F}$-identification function $V$ are linked via the derivative ([1], [21], [22])

$$\frac{\partial L(z,y)}{\partial z} = h(z)V(z, y) \tag{2.8}$$

where $h$ is a nonnegative function and $V$ is an oriented function. The function $V$ is an oriented strict $\mathcal{F}$-identification function if it is a strict $\mathcal{F}$-identification function for the functional $T(F)$ and, moreover $\mathbb{E}_F[V(z, \underline{y})] > 0$ if and only if $z > T(F)\ \forall\ F \in \mathcal{F}, z \in D \subseteq \mathbb{R}$. ([21], [23]).

Common identification function examples include $V_{\text{median}}$ (eq. (B.6)) for the median and $V_{\text{mean}}$ (eq. (B.7)) for the mean. In practice, the empirical average identification function $\bar{V}$ is computed on a test set

$$\bar{V}(\boldsymbol{z}, \boldsymbol{y}) = (1/n)\sum_{i=1}^{n} V(z_i, y_i) \tag{2.9}$$

Commonly used empirical averages include $\bar{V}_{\text{median}}$ (eq. (B.8)) and $\bar{V}_{\text{mean}}$ (eq. (B.9)), the latter also known as mean error (ME). Adequate predictions for the functional $T$ should yield $\bar{V} \approx 0$ (subject to scaling of the predicted variable). For example, a median prediction is adequate if roughly half the predictions exceed their corresponding observations.

Finally, it is possible for multiple models to be adequate for the same functional (e.g., both identifying the mean) yet exhibit different average losses. In such cases, the model with the lower average loss (e.g. lower MSE for mean predictions) is still preferred, as it is more accurate even though both are calibrated.

#### *2.1.5 Prespecifying the loss function in model estimation and prediction evaluation*

Theoretical work establishes that the loss function used for ranking must be disclosed a priori, as model rankings depend critically on this choice ([1], [24]). This dependency can be illustrated with an example. Consider Model #1, trained to predict conditional medians using the absolute error (AE) loss, and Model #2, trained to predict conditional means using the squared error (SE) loss. If the dependent variable follows an asymmetric distribution, their relative performance reverses depending on the loss function, since Model #1 will outperform Model #2 under AE loss, while the opposite holds under SE loss [10]. This follows directly from the equivalence between the consistency of an *M*-estimator (used in training) and evaluation with a strictly consistent loss function.

### 2.2 Consistent extremum and *M*-estimators

The general class of extremum estimators was formalized by [7] [25] and [26] and. Consider a loss function $L(\boldsymbol{\theta}, \boldsymbol{y})$ that quantifies the discrepancy for a parameter vector $\boldsymbol{\theta}$ given a realization $\boldsymbol{y} = (y_1, \dots, y_n)^{\mathrm{T}}$. Assuming the true parameter $\boldsymbol{\theta}_0$ lies in the parameter space $\boldsymbol{\Theta}$, an extremum estimator $\widehat{\boldsymbol{\theta}}(\boldsymbol{y})$ for $\boldsymbol{\theta}_0$ is defined as:

$$\widehat{\boldsymbol{\theta}}(\underline{\boldsymbol{y}}) := \operatorname*{argmin}_{\boldsymbol{\theta} \in \boldsymbol{\Theta}} L(\boldsymbol{\theta}, \underline{\boldsymbol{y}}) \tag{2.10}$$

The properties of $\widehat{\boldsymbol{\theta}}(\underline{\boldsymbol{y}})$ depend on the sample size $n$ through the realizations $\boldsymbol{y}$. While $L$ can take general forms, the additive case has been extensively studied. When $L$ is additive, the resulting estimators are called *M*-estimators ([27], [28]). For a one-dimensional parameter $\theta_0$, this takes the form:

$$\hat{\theta}(\underline{\boldsymbol{y}}) := \operatorname*{argmin}_{\theta \in \Theta} \textstyle\sum_{i=1}^{n} L(\theta, \underline{y_i}) \tag{2.11}$$

A canonical example is the SE loss, which yields the *M*-estimator for the mean:

$$\hat{\mu}(\underline{\boldsymbol{y}}) := \operatorname*{argmin}_{\theta \in \Theta} \textstyle\sum_{i=1}^{n} (\theta - \underline{y_i})^2 , \Theta \subseteq \mathbb{R} \tag{2.12}$$

Consistency is a key property for an estimator. An estimator $\widehat{\boldsymbol{\theta}}$ is consistent if it converges in probability to the true parameter value $\boldsymbol{\theta}_0$, i.e. $\widehat{\boldsymbol{\theta}}(\underline{\boldsymbol{y}}) \xrightarrow{P} \boldsymbol{\theta}_0$. For instance, $\hat{\mu}(\underline{\boldsymbol{y}})$ is consistent for $\mathbb{E}_{F_{\underline{y}}}[\underline{y}]$, where $F_{\underline{y}}$ is the cumulative distribution function (CDF) of $\underline{y}$, and $\mathbb{E}_{F_{\underline{y}}}[\underline{y}]$ is the expectation of $\underline{y}$.

A fundamental connection exists between the two distinct procedures of estimation and prediction evaluation for the case of $M$-estimators. Under general conditions, an $M$-estimator is consistent for the parameter $\theta_0$ if and only if the loss function $L$ in eq. (2.11) is strictly consistent for the functional $T$ corresponding to that parameter [10].

### 2.3 The original index of agreement loss function

The index of agreement loss function admits the following vectorized representation:

$$L_W(\boldsymbol{z},\boldsymbol{y}) = \|\boldsymbol{z}-\boldsymbol{y}\|_2^2/\|(|\boldsymbol{z}-\mathbf{1}\mu(\boldsymbol{y})| + |\mathbf{1}\mu(\boldsymbol{y})-\boldsymbol{y}|)\|_2^2 \tag{2.13}$$

where $\boldsymbol{z} = (z_1, \dots, z_n)^{\mathrm{T}}$ represents the vector of model predictions, $\boldsymbol{y} = (y_1, \dots, y_n)^{\mathrm{T}}$ represents the vector of observed values, $\|\boldsymbol{x}\|_2$ is the Euclidean norm (eq. (A.9)), $|\boldsymbol{x}|$ indicates the element-wise absolute value (eq. (A.4)) and $\mathbf{1}$ denotes the all-ones vector defined in eq. (A.3). $L_W$ is defined when $\boldsymbol{y}$ is non-constant (i.e., has at least one element different from the others) or $\boldsymbol{z} \neq \boldsymbol{y}$ (where $\neq$ indicates pairwise inequality for at least one vector element; see eq. (A.6)).

This vectorized form simplifies theoretical analysis. For instance, Proof D.1 in Appendix D establishes that $L_W$ is bounded:

$$0 \leq L_W(\boldsymbol{z},\boldsymbol{y}) \leq 1 \tag{2.14}$$

As noted in the Introduction, $L_W$ is negatively oriented. Furthermore, $L_W(\boldsymbol{z},\boldsymbol{y}) = 0$, if and only if $\boldsymbol{z} = \boldsymbol{y}$ and $\boldsymbol{y}$ is non-constant.

## 3. The index of agreement

This section examines the original index of agreement and introduces theoretically consistent variants developed from its foundational construction principles. Formal proofs for the presented results are given in Appendix D.

### 3.1 The index of agreement loss function

This section details key properties of the original index of agreement loss function $L_W$.

#### *3.1.1 Translation invariance*

The loss function $L_W$ is translation invariant, meaning that shifting predictions and realizations by a constant $c \in \mathbb{R}$ its value does not change (see Proof D.2):

$$L_W(\boldsymbol{z} + c\mathbf{1}, \boldsymbol{y} + c\mathbf{1}) = L_W(\boldsymbol{z}, \boldsymbol{y}), c \in \mathbb{R} \tag{3.1}$$

### *3.1.2 Scale invariance*

The loss function $L_W$ exhibits scale invariance, meaning its value remains unchanged when both predictions and observations are scaled by a non-zero-constant $c \in \mathbb{R}\backslash\{0\}$:

$$L_W(c\boldsymbol{z}, c\boldsymbol{y}) = L_W(\boldsymbol{z}, \boldsymbol{y}), c \in \mathbb{R}\backslash\{0\} \tag{3.2}$$

This property follows directly from the definition in eq. (2.13). Therefore, its behavior remains unchanged across different measurement units.

### *3.1.3 Extremum estimation*

To analyze $L_W$ as an extremum estimator for a scalar parameter $\theta$, we minimize it for a constant vector $\theta\mathbf{1}$, given non-constant observations $\boldsymbol{y}$:

$$\hat{\theta}(\underline{\boldsymbol{y}}) := \underset{\theta \in \Theta}{\operatorname{argmin}}\, L_W(\theta\mathbf{1}, \underline{\boldsymbol{y}}) \tag{3.3}$$

As shown in Proof D.3, this minimization yields two distinct estimators:

$$\hat{\theta}(\underline{\boldsymbol{y}}) = \mu(\underline{\boldsymbol{y}}) \pm \sigma(\underline{\boldsymbol{y}}) \tag{3.4}$$

Figure 1 illustrates $L_W(\theta\mathbf{1}, \boldsymbol{y})$ as a function of $\theta$ for a realization $\boldsymbol{y}$ of $\underline{\boldsymbol{y}}$, indicating the points where it attains its minimum. The minimum value of $L_W(\theta\mathbf{1}, \underline{\boldsymbol{y}})$ is $(\sigma(\boldsymbol{y}))/(\sigma(\boldsymbol{y}) + \mathrm{mad}(\boldsymbol{y}))$, where $\mathrm{mad}(\boldsymbol{y})$ is the mean absolute deviation defined in eq. (A.20).

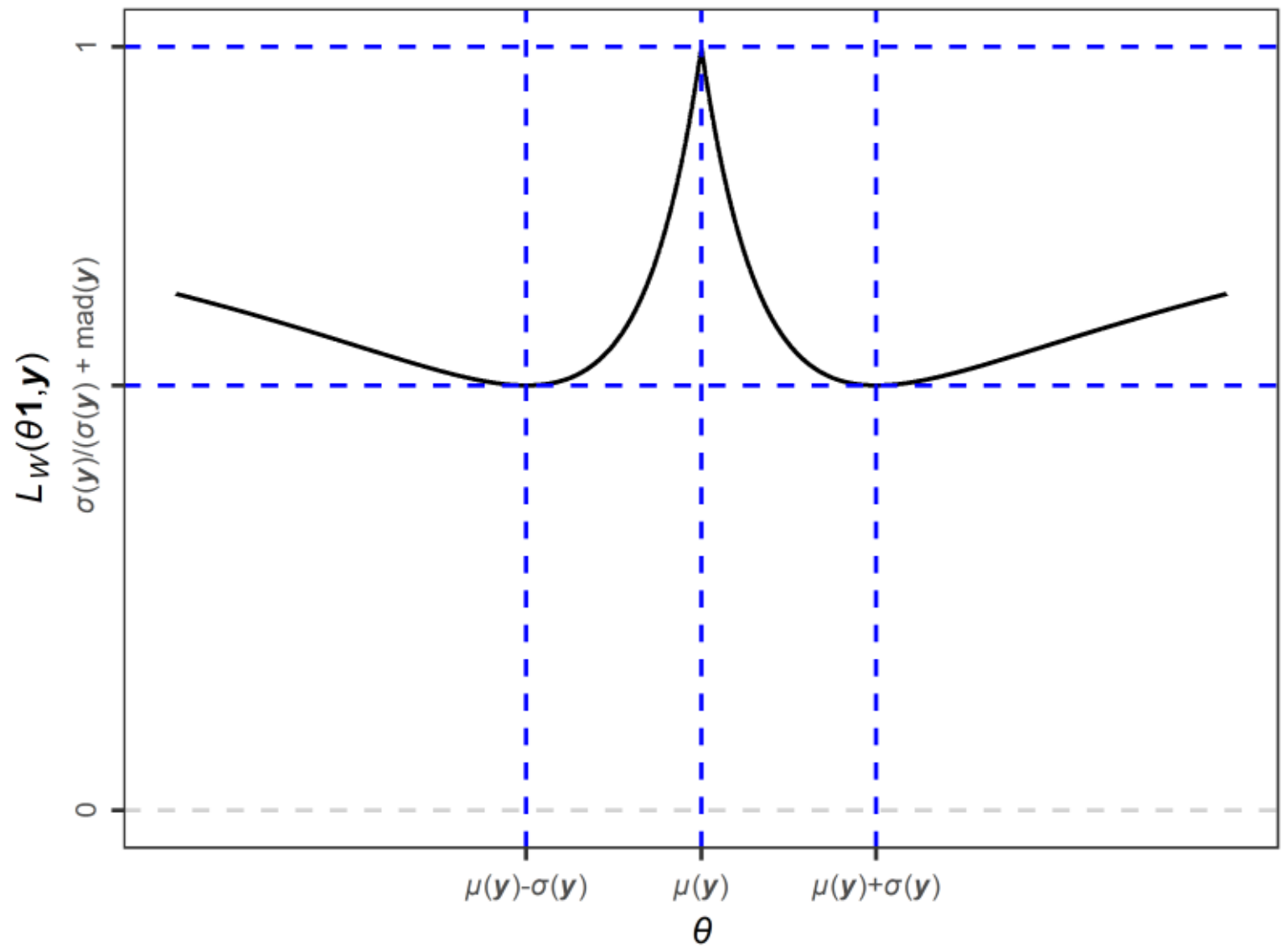

Figure 1. Illustration of $L_W(\theta\mathbf{1}, \boldsymbol{y})$ function.

The non-uniqueness precludes consistency, as the estimator does not converge to a single parameter value. To resolve this, we define two modified estimators, i.e. the upper-

bound estimator

$$\hat{\theta}_+(\underline{\boldsymbol{y}}) = \max\{\underset{\theta\in\Theta}{\operatorname{argmin}} L_W(\theta\mathbf{1}, \underline{\boldsymbol{y}})\} = \mu(\underline{\boldsymbol{y}}) + \sigma(\underline{\boldsymbol{y}}) \tag{3.5}$$

and the lower-bound estimator

$$\hat{\theta}_-(\underline{\boldsymbol{y}}) = \min\{\underset{\theta\in\Theta}{\operatorname{argmin}} L_W(\theta\mathbf{1}, \underline{\boldsymbol{y}})\} = \mu(\underline{\boldsymbol{y}}) - \sigma(\underline{\boldsymbol{y}}) \tag{3.6}$$

Proof D.4 establishes that, under independent identically distributed (IID) variables assumption, $\hat{\theta}_+(\underline{\boldsymbol{y}})$ and $\hat{\theta}_-(\underline{\boldsymbol{y}})$ are consistent for $\mathbb{E}_F[\underline{y}] + \mathbb{V}_F^{1/2}[\underline{y}]$ and $\mathbb{E}_F[\underline{y}] - \mathbb{V}_F^{1/2}[\underline{y}]$, respectively, where $\mathbb{V}_F[\underline{y}]$ is the variance of $\underline{y}$.

### *3.1.4 Estimation of a linear model*

An open problem is the estimation of the true parameters $\boldsymbol{\theta}_0 = (a_0, b_0) \in \mathbb{R}^2$ for the linear model $\boldsymbol{z} = a\boldsymbol{x} + b\mathbf{1}$, using the $L_W$ loss. The estimator

$$\widehat{\boldsymbol{\theta}}(\underline{\boldsymbol{x}}, \underline{\boldsymbol{y}}) := \underset{\boldsymbol{\theta}\in\mathbb{R}^2}{\operatorname{argmin}} L_W(a\underline{\boldsymbol{x}} + b\mathbf{1}, \underline{\boldsymbol{y}}) \tag{3.7}$$

lacks a known closed-form solution.

## 3.2 The new agreement loss function

The original index of agreement (eq. (2.13)) is fundamentally a ratio of two squared Euclidean norms. While the numerator $\|\boldsymbol{z} - \boldsymbol{y}\|_2^2$ directly measures the magnitude of prediction errors, the denominator involves the norm of the composite vector $|\boldsymbol{z} - \mathbf{1}\mu(\boldsymbol{y})| + |\mathbf{1}\mu(\boldsymbol{y}) - \boldsymbol{y}|$. This construction, which applies element-wise absolute values, complicates geometric interpretation.

To address this, we propose an alternative loss function $L_{\mathrm{NR}_2}$ based on the Euclidean distances between the vectors $\boldsymbol{z}, \boldsymbol{y}$ and the mean vector $\mathbf{1}\mu(\boldsymbol{y})$. These vectors form a triangle in $\mathbb{R}^n$, satisfying the triangle inequality in eq. (A.31)

$$0 \le \|\boldsymbol{z} - \boldsymbol{y}\|_2 \le \|\boldsymbol{z} - \mathbf{1}\mu(\boldsymbol{y})\|_2 + \|\mathbf{1}\mu(\boldsymbol{y}) - \boldsymbol{y}\|_2 \tag{3.8}$$

Leveraging this inequality, we define:

$$L_{\mathrm{NR}_2}(\boldsymbol{z}, \boldsymbol{y}) := \|\boldsymbol{z} - \boldsymbol{y}\|_2^2 / (\|\boldsymbol{z} - \mathbf{1}\mu(\boldsymbol{y})\|_2 + \|\mathbf{1}\mu(\boldsymbol{y}) - \boldsymbol{y}\|_2)^2 \tag{3.9}$$

where the subscript NR stands for norm ratio. $L_{\mathrm{NR}_2}$ is defined when $\boldsymbol{y}$ is non-constant or $\boldsymbol{z} \neq \boldsymbol{y}$. An equivalent expression is:

$$L_{\mathrm{NR}_2}(\boldsymbol{z}, \boldsymbol{y}) = (\textstyle\sum_{i=1}^n (z_i - y_i)^2) / (\sqrt{\sum_{i=1}^n (z_i - \mu(\boldsymbol{y}))^2} + \sqrt{\sum_{i=1}^n (\mu(\boldsymbol{y}) - y_i)^2})^2 \tag{3.10}$$

$L_{\mathrm{NR}_2}$ has the following geometric and theoretical advantages:

(i) (Intuitive geometric foundation): Unlike the original index in eq. (1.1), where the denominator sums element-wise absolute deviations before applying the Euclidean norm, $L_{\mathrm{NR}_2}$ operates directly on the vectors $\boldsymbol{z} - \mathbf{1}\mu(\boldsymbol{y})$ and $\mathbf{1}\mu(\boldsymbol{y}) - \boldsymbol{y}$. This aligns with the Euclidean triangle inequality in eq. (3.8), treating prediction errors as vector displacements in $\mathbb{R}^n$. The denominator represents the total path length from $\boldsymbol{z}$ to $\boldsymbol{y}$ via $\mathbf{1}\mu(\boldsymbol{y})$, while the numerator is the direct displacement.

(ii) (Theoretical consistency): The element-wise absolute values in the original denominator disrupt vector-space properties. By eliminating these, $L_{\mathrm{NR}_2}$ preserves translation and scale invariance (inherited from $L_W$, see Sections 3.2.3 and 3.2.2) and maintains compatibility with norm axioms avoiding ad hoc element-wise operations. This enables cleaner analysis (e.g. closed-form solutions for linear models; see Section 3.2.5).

(iii) (Interpretability): $L_{\mathrm{NR}_2}$ quantifies the inefficiency of predictions relative to a "baseline path" through $\mathbf{1}\mu(\boldsymbol{y})$.

### *3.2.1 Bounds*

From eqs. (3.8) and (3.9), it follows immediately that:

$$0 \leq L_{\mathrm{NR}_2}(\boldsymbol{z}, \boldsymbol{y}) \leq 1 \tag{3.11}$$

Furthermore, $L_{\mathrm{NR}_2}(\boldsymbol{z}, \boldsymbol{y}) = 0$, if and only if $\boldsymbol{z} = \boldsymbol{y}$ and $\boldsymbol{y}$ is non-constant. Applying eq. (A.32) shows that $L_{\mathrm{NR}_2}$ attains its upper bound of 1 in the following cases:

(i) When $\boldsymbol{y} = \mathbf{1}\mu(\boldsymbol{y})$ and $\boldsymbol{z} \neq \boldsymbol{y}$.

(ii) When $\boldsymbol{z} = \mathbf{1}\mu(\boldsymbol{y})$ and $\boldsymbol{y}$ is not constant.

(iii) When $\boldsymbol{z} - \mathbf{1}\mu(\boldsymbol{y}) = a(\mathbf{1}\mu(\boldsymbol{y}) - \boldsymbol{y})$ for some $a > 0$ and $\boldsymbol{y}$ is not constant.

### *3.2.2 Translation invariance*

Analogous to $L_W$, the loss function $L_{\mathrm{NR}_2}$ is translation invariant (see Proof D.5):

$$L_{\mathrm{NR}_2}(\boldsymbol{z} + c\mathbf{1}, \boldsymbol{y} + c\mathbf{1}) = L_{\mathrm{NR}_2}(\boldsymbol{z}, \boldsymbol{y}), c \in \mathbb{R} \tag{3.12}$$

### *3.2.3 Scale invariance*

The scale invariance of $L_{\mathrm{NR}_2}$

$$L_{\mathrm{NR}_2}(c\boldsymbol{z}, c\boldsymbol{y}) = L_{\mathrm{NR}_2}(\boldsymbol{z}, \boldsymbol{y}), c \in \mathbb{R}\backslash\{0\} \tag{3.13}$$

follows directly from its definition in eq. (3.9).

### 3.2.4 *Extremum estimation*

To analyze $L_{\mathrm{NR}_2}$ as an extremum estimator for a scalar parameter $\theta$, we minimize it for a constant vector $\theta\mathbf{1}$, given non-constant observations $\boldsymbol{y}$:

$$\hat{\theta}(\underline{\boldsymbol{y}}) := \operatorname*{argmin}_{\theta\in\Theta} L_{\mathrm{NR}_2}(\theta\mathbf{1}, \underline{\boldsymbol{y}}) \tag{3.14}$$

As shown in Proof D.6, this minimization yields two distinct estimators:

$$\hat{\theta}(\underline{\boldsymbol{y}}) = \mu(\underline{\boldsymbol{y}}) \pm \sigma(\underline{\boldsymbol{y}}) \tag{3.15}$$

Figure 2 illustrates $L_{\mathrm{NR}_2}(\theta\mathbf{1}, \boldsymbol{y})$ as a function of $\theta$ for a realization $\boldsymbol{y}$ of $\underline{\boldsymbol{y}}$, indicating the points where it attains its minimum. The minimum value of $L_{\mathrm{NR}_2}(\theta\mathbf{1}, \underline{\boldsymbol{y}})$ is $1/2$.

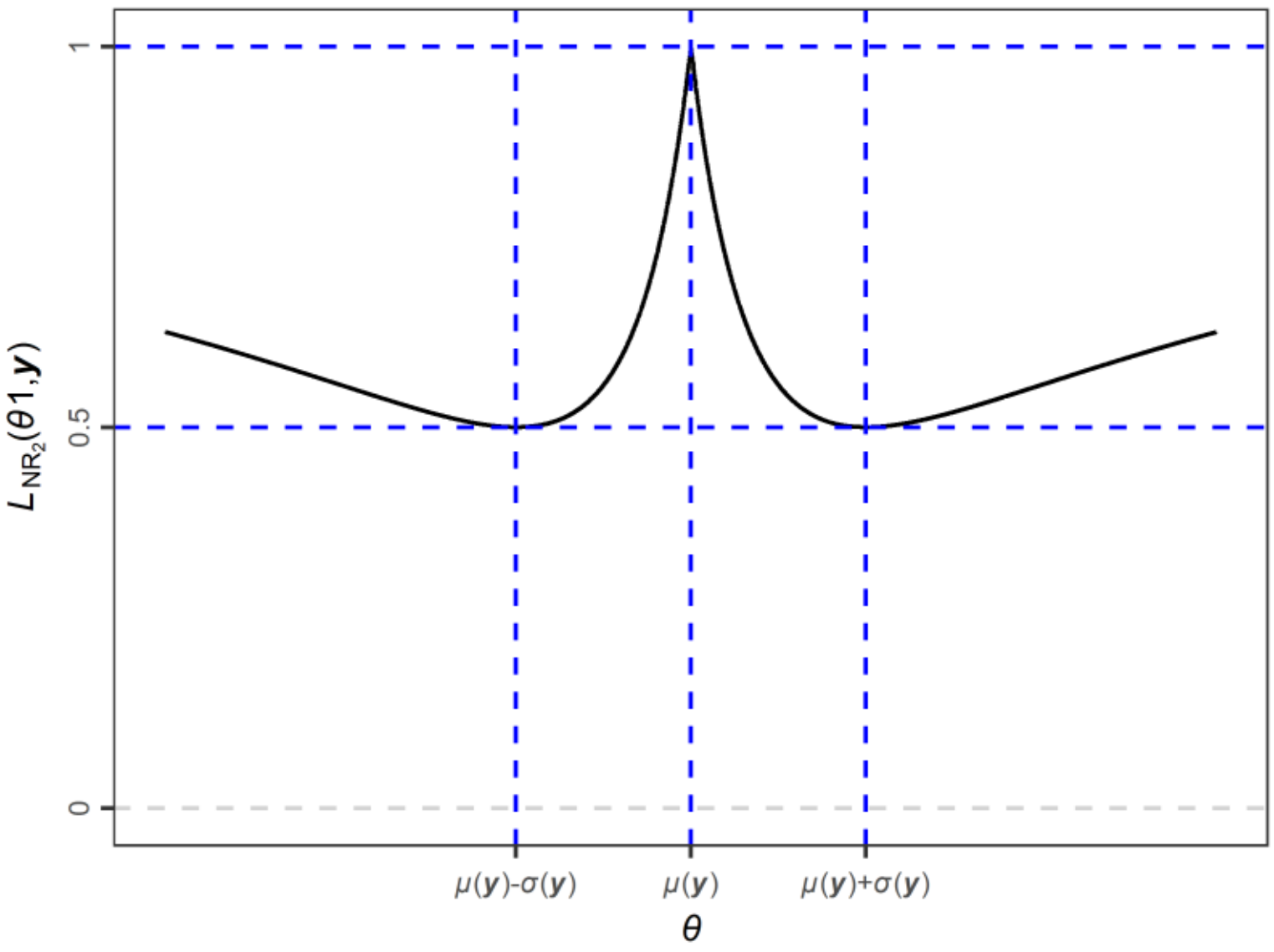

Figure 2. Illustration of $L_{\mathrm{NR}_2}(\theta\mathbf{1}, \boldsymbol{y})$ function.

The non-uniqueness precludes consistency, as the estimator does not converge to a single parameter value. To resolve this, we define the upper-bound estimator

$$\hat{\theta}_{+}(\underline{\boldsymbol{y}}) = \max\{\operatorname*{argmin}_{\theta\in\Theta} L_{\mathrm{NR}_2}(\theta\mathbf{1}, \underline{\boldsymbol{y}})\} = \mu(\underline{\boldsymbol{y}}) + \sigma(\underline{\boldsymbol{y}}) \tag{3.16}$$

and the lower-bound estimator

$$\hat{\theta}_{-}(\underline{\boldsymbol{y}}) = \min\{\operatorname*{argmin}_{\theta\in\Theta} L_{\mathrm{NR}_2}(\theta\mathbf{1}, \underline{\boldsymbol{y}})\} = \mu(\underline{\boldsymbol{y}}) - \sigma(\underline{\boldsymbol{y}}) \tag{3.17}$$

Under the IID assumption, Proof D.4 establishes that $\hat{\theta}_{+}(\underline{\boldsymbol{y}})$ and $\hat{\theta}_{-}(\underline{\boldsymbol{y}})$ are consistent for $\mathbb{E}_F[\underline{y}] + \mathbb{V}_F^{1/2}[\underline{y}]$ and $\mathbb{E}_F[\underline{y}] - \mathbb{V}_F^{1/2}[\underline{y}]$, respectively.

### 3.2.5 *Estimation of a linear model*

For the linear model $\boldsymbol{z} = a\boldsymbol{x} + b\mathbf{1}$, we estimate the true parameters $\boldsymbol{\theta}_0 = (a_0, b_0) \in \mathbb{R}^2$, using the $L_{\mathrm{NR}_2}$ loss, via the extremum estimator $\widehat{\boldsymbol{\theta}} = (\hat{a}, \hat{b})$

$$\widehat{\boldsymbol{\theta}}(\underline{\boldsymbol{x}}, \underline{\boldsymbol{y}}) := \underset{\boldsymbol{\theta} \in \mathbb{R}^2}{\operatorname{argmin}} L_{\mathrm{NR}_2}(a\underline{\boldsymbol{x}} + b\mathbf{1}, \underline{\boldsymbol{y}}) \tag{3.18}$$

As shown in Proof D.7 (eqs. (D.123) and (D.124)), the solution is:

$$\widehat{\boldsymbol{\theta}}(\underline{\boldsymbol{x}}, \underline{\boldsymbol{y}}) = (\operatorname{sign}(\rho(\underline{\boldsymbol{x}}, \underline{\boldsymbol{y}}))(||\underline{\boldsymbol{y}}_c||_2/||\underline{\boldsymbol{x}}_c||_2), \mu(\underline{\boldsymbol{y}}) - (\operatorname{sign}(\rho(\underline{\boldsymbol{x}}, \underline{\boldsymbol{y}}))\mu(\underline{\boldsymbol{x}})||\underline{\boldsymbol{y}}_c||_2)/ ||\underline{\boldsymbol{x}}_c||_2) \tag{3.19}$$

where $\boldsymbol{x}_c$ and $\boldsymbol{y}_c$ are centered vectors defined in eq. (A.23), $\rho(\boldsymbol{x}, \boldsymbol{y})$ is the Pearson correlation defined in eq. (A.28) and $\operatorname{sign}(x)$ is the scalar sign function, defined in eq. (A.16). At this point, $L_{\mathrm{NR}_2}$ attains its minimum value (see eq. (D.125)):

$$L_{\mathrm{NR}_2}(\hat{a}\boldsymbol{x} + \hat{b}\mathbf{1}, \boldsymbol{y}) = (1 - |\rho(\boldsymbol{x}, \boldsymbol{y})|)/2 \tag{3.20}$$

When a correctly specified linear model is trained using the SE loss, its predictions correspond to the conditional mean $\mathbb{E}_{F_{\underline{y}|\underline{\boldsymbol{x}}}}[\underline{y}|\underline{\boldsymbol{x}}]$. This occurs because the SE-derived estimator is an $M$-estimator, and the SE loss is strictly consistent for the mean (see Section 2.1.2), i.e., it is uniquely minimized in expectation by the mean prediction [10]. While such results are well-established for $M$-estimators, their extension to the broader class of extremum estimators remains an open problem. For completeness, we also present the standard least-squares regression parameter estimates:

$$\widehat{\boldsymbol{\theta}}(\underline{\boldsymbol{x}}, \underline{\boldsymbol{y}}) = (\rho(\underline{\boldsymbol{x}}, \underline{\boldsymbol{y}}))(||\underline{\boldsymbol{y}}_c||_2/||\underline{\boldsymbol{x}}_c||_2), \mu(\underline{\boldsymbol{y}}) - (\rho(\underline{\boldsymbol{x}}, \underline{\boldsymbol{y}})\mu(\underline{\boldsymbol{x}})||\underline{\boldsymbol{y}}_c||_2)/||\underline{\boldsymbol{x}}_c||_2) \tag{3.21}$$

## 3.3 Generalization of the new index of agreement

Legates and McCabe Jr [8] proposed the following loss function as an improvement of the original index of agreement

$$L_{\mathrm{LMC}}(\boldsymbol{z}, \boldsymbol{y}) := \textstyle\sum_{i=1}^{n} |z_i - y_i| \,/\, \sum_{i=1}^{n} (|z_i - f(\boldsymbol{y})| + |f(\boldsymbol{y}) - y_i|) \tag{3.22}$$

where $f(\boldsymbol{y})$ is a benchmark method (e.g. climatology or persistence). $L_{\mathrm{LMC}}$ is the equal-weight special case of a weighted variant of the index of agreement introduced by [29]. It admits the vectorized formulation:

$$L_{\mathrm{LMC}}(\boldsymbol{z}, \boldsymbol{y}) = \|\boldsymbol{z} - \boldsymbol{y}\|_1 / \|(|\boldsymbol{z} - \mathbf{1}f(\boldsymbol{y})| + |\mathbf{1}f(\boldsymbol{y}) - \boldsymbol{y}|)\|_1 \tag{3.23}$$

where $\|\boldsymbol{x}\|_1$ is the Manhattan norm, defined in eq. (A.8).

Krause et al. [9] later introduced a generalization of the form

$$L_{\mathrm{KBB}}(\boldsymbol{z}, \boldsymbol{y}) := \textstyle\sum_{i=1}^{n} |z_i - y_i|^p \,/\, \sum_{i=1}^{n} (|z_i - \mu(\boldsymbol{y})| + |\mu(\boldsymbol{y}) - y_i|)^p \,, p = 1,2, \ldots \tag{3.24}$$

This admits the vectorized formulation

$$L_{\mathrm{KBB}}(\boldsymbol{z},\boldsymbol{y}) = \|\boldsymbol{z}-\boldsymbol{y}\|_p^p / \|(|\boldsymbol{z}-\mathbf{1}\mu(\boldsymbol{y})| + |\mathbf{1}\mu(\boldsymbol{y})-\boldsymbol{y}|)\|_p^p, p = 1,2,\dots \quad (3.25)$$

where $\|\boldsymbol{x}\|_p$ is the $L_p$-norm, defined in eq. (A.10), and the Chebyshev norm as the special case for $p = \infty$, defined in eq. (A.11). $L_{\mathrm{KBB}}$ is another equal-weight special case of the weighted variant introduced by [29], for $p = 1, 2, \dots$. However, this definition shares the theoretical limitations of $L_W$ (see Section 3.2). Furthermore, it employs the mean climatology $\mathbf{1}\mu(\boldsymbol{y})$ as the benchmark, which aligns naturally with Euclidean norms but not necessarily with other $L_p$-norms. This is less flexible than the definition of $L_{\mathrm{LMC}}$, which allows for a tailored climatology $f(\boldsymbol{y})$. For instance, the Manhattan norm $\|\boldsymbol{x}\|_1$ corresponds more naturally to the median climatology $\mathbf{1}\mathrm{median}(\boldsymbol{y})$ [1].

Analogous to $L_{\mathrm{NR}_2}$, we propose the loss function $L_{\mathrm{NR}_p}$ (defined for non-constant $\boldsymbol{y}$ or $\boldsymbol{z} \neq \boldsymbol{y}$)

$$L_{\mathrm{NR}_p}(\boldsymbol{z},\boldsymbol{y}) := \|\boldsymbol{z}-\boldsymbol{y}\|_p^p / \left(\left\|\boldsymbol{z}-\mathbf{1}L_{p-\mathrm{mean}}(\boldsymbol{y})\right\|_p + \left\|\mathbf{1}L_{p-\mathrm{mean}}(\boldsymbol{y})-\boldsymbol{y}\right\|_p\right)^p \quad (3.26)$$

where $L_{p-\mathrm{mean}}(\boldsymbol{y})$ is the $L_p$ mean of $\boldsymbol{y}$ [30]. Examples of $L_{p-\mathrm{mean}}$ are the sample median, which equals the $L_1$ mean, and the sample mean, which equals the $L_2$ mean. The proposed $L_{\mathrm{NR}_p}$ retains the advantages of $L_{\mathrm{NR}_2}$ (see Section 3.2), including its intuitive geometric foundation, theoretical consistency and interpretability, while adapting the benchmark method to the chosen norm. For $p = 1$ and setting $f(\boldsymbol{y}) = L_{p-\mathrm{mean}}(\boldsymbol{y})$, the $L_{\mathrm{LMC}}$ loss is a special case of the $L_{\mathrm{NR}_p}$ loss. Although a complete theoretical analysis of $L_{\mathrm{NR}_p}$ is left for subsequent work, the following bound follows directly from its construction:

$$0 \leq L_{\mathrm{NR}_p}(\boldsymbol{z},\boldsymbol{y}) \leq 1 \quad (3.27)$$

## 4. Applications

This section presents applications designed to explore the implications of our theoretical insights. We analyze both simulated data and real-world case studies using hydrologic models. All computations were performed in `R` (version 4.5.1, [31]) within the `RStudio` integrated development environment (version 2025.09.0+387). Supplementary materials provide reproducible code and datasets for these analyses, and the specific `R` packages used are listed in Appendix E.

### 4.1 Predicting a Gaussian random variable

In the first example, we simulate data from a Gaussian random variable $\underline{y} \sim N(\mu, \sigma)$ (whose probability density function (PDF) is defined in eq. (C.1)). A sample of $n = 1\,000$ realizations was simulated with true parameters $\mu = 0$ and $\sigma = 1$. Using the first 500 samples as the training set, we computed three climatology-based predictive models:

(i) Model #1: $z = \mu(\boldsymbol{y}_{1:500})$.

(ii) Model #2: $z = \mu(\boldsymbol{y}_{1:500}) - \sigma(\boldsymbol{y}_{1:500})$.

(iii) Model #3: $z = \mu(\boldsymbol{y}_{1:500}) + \sigma(\boldsymbol{y}_{1:500})$.

These three models were then used to forecast the remaining 500 samples (the test set), as illustrated in Figure 3.

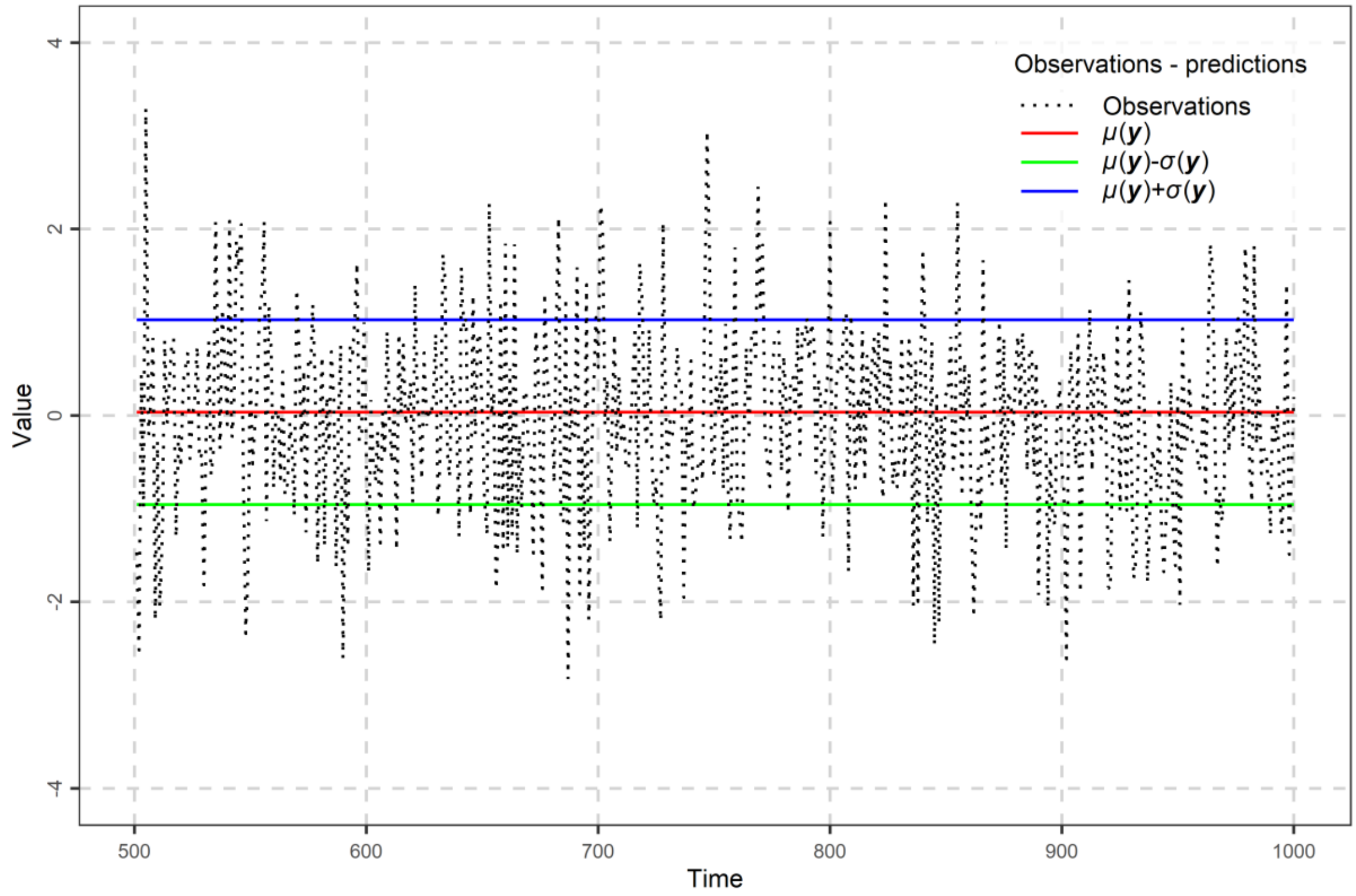

Figure 3. 500 observations from the test set of a random variable $\underline{y} \sim N(0,1)$. The values $\mu(\boldsymbol{y}), \mu(\boldsymbol{y}) - \sigma(\boldsymbol{y}), \mu(\boldsymbol{y}) + \sigma(\boldsymbol{y})$ were computed from the 500 obsevations in the training set and are used as constant predictions in the test set.

We evaluate the performance of these three models on the test set in Table 1. To align with the negative orientation of the other metrics (MSE, $L_{\mathrm{NR}_2}$ and $L_W$), we use $1 - \mathrm{NSE}(\boldsymbol{z}, \boldsymbol{y})$, instead of the conventional NSE. This transformed metric is bounded below by 0.

Table 2. Performance of models on the test set. Predictions $\boldsymbol{z}$ are the three climatologies computed from the training set (models #1, #2 and #3), while observations $\boldsymbol{y} = \boldsymbol{y}_{501:1000}$ are from the test set.

| Metric | Model #1 | Model #2 | Model #3 |
|---|---|---|---|
| $\mathrm{MSE}(\boldsymbol{z}, \boldsymbol{y})$ | 1.020768 | 1.858657 | 2.146621 |
| $1 - \mathrm{NSE}(\boldsymbol{z}, \boldsymbol{y})$ | 1.005198 | 1.830307 | 2.113878 |
| $L_{\mathrm{NR}_2}(\boldsymbol{z}, \boldsymbol{y})$ | 0.8745487 | 0.5010791 | 0.5003633 |
| $L_W(\boldsymbol{z}, \boldsymbol{y})$ | 0.8979654 | 0.5590595 | 0.5583535 |

Due to random sampling, the results on the test set deviate slightly from the theoretical values expected from the training set (omitted for brevity). However, the sample is sufficiently large to confirm the theoretical findings from Section 3. As noted in Section 2.1.3, using MSE or $1 - \mathrm{NSE}$ does not alter the model rankings. Model #1 outperforms the others with respect to MSE, as expected since the MSE is based on the $L_{\mathrm{SE}}$ loss, strictly consistent for the mean. Models #2 and #3 perform similarly under MSE, a result subject to random variation but expected given the symmetry of both the $L_{\mathrm{SE}}$ loss and the Gaussian distribution.

In contrast, the rankings are completely reversed when comparing models using the $L_{\mathrm{NR}_2}$ and $L_W$ losses. These results still align with the theory developed in Section 3. Models #2 and #3 demonstrate similar performance and significantly outperform model #1 under both metrics. The value of $L_{\mathrm{NR}_2}$ approaches its theoretical minimum of 0.50 (subject to randomness), confirming the finding illustrated in Figure 2. Similarly, the values of $L_W$ are near their theoretical minimum $((\sigma(\boldsymbol{y}))/(\sigma(\boldsymbol{y}) + \mathrm{mad}(\boldsymbol{y})))$, as shown in Figure 1.

These results may appear paradoxical at first glance. Based on Figure 3 and prior experience with the $L_W$ loss, one might expect model #1 to outperform the others on all metrics. However, the superior performance of models #2 and #3 under the $L_{\mathrm{NR}_2}$ and $L_W$ losses is explained in Sections 3.1.3 and 3.2.4. These sections establish that the climatologies $\mu(\underline{y}) \pm \sigma(\underline{y})$ are the minimizers for these specific loss functions, when fitting to a sample from a probability distribution.

## 4.2 Predicting data from a linear model with error

In the second example, we simulate data from a Gamma-distributed predictor random variable $\underline{x} \sim \mathrm{Gamma}(\theta, a)$ (whose PDF is defined in eq. (C.2)) and a log-normal error random variable $\underline{\varepsilon} \sim \mathrm{Lognormal}(\mu, \sigma)$ (whose PDF is defined in eq. (C.3)). We generate $n =$ 4 000 realizations using the true parameters $\theta = 1.8$, $a = 0.4$, $\mu = 0$ and $\sigma = 2$. We then use these simulations to generate 4 000 realizations of $\underline{y}$, from the linear model $\underline{y} = a_0 +$ $a_1\underline{x} + \underline{\varepsilon}$, with a fixed intercept $a_0 = 2.1$. We examine three cases for the slope coefficient $a_1$: $a_1 = 0.6$ (Figure 4a), $a_1 = 6.0$ (Figure 4b) and $a_1 = 20.0$ (Figure 4c). We intentionally introduced highly asymmetric variables in this simulation to ensure the results are sufficiently general, despite their empirical nature.

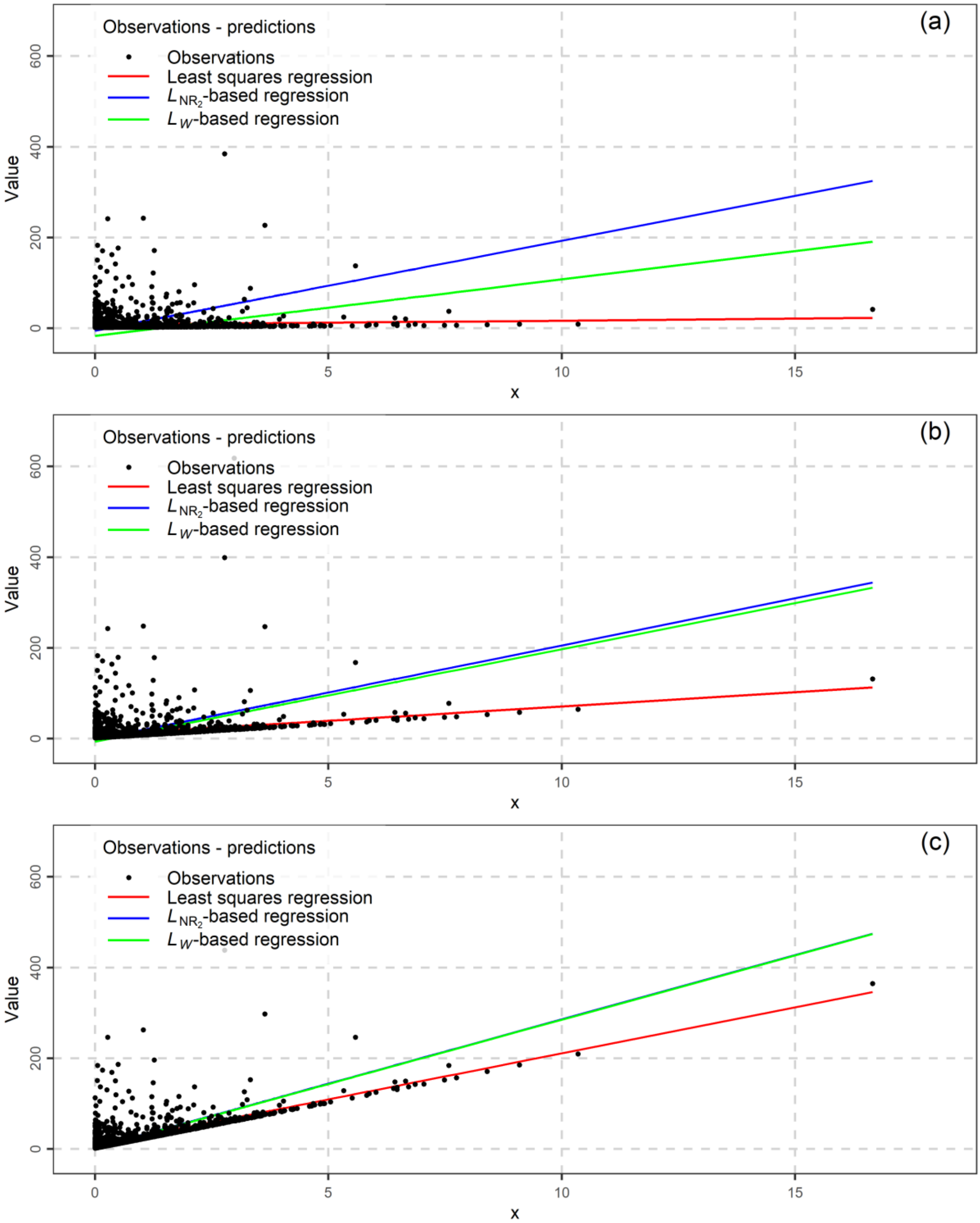

Figure 4. Predictions of linear models fitted using different loss functions. Black dots represent 2 000 test observations of $\underline{y} = 2.1 + a_1\underline{x} + \underline{\varepsilon}$, where $\underline{x} \sim \text{Gamma}(1.8, 0.4)$ and $\underline{\varepsilon} \sim \text{Lognormal}(0, 2)$. Results are shown for three slope values: (a) $a_1 = 0.6$, (b) $a_1 = 6.0$, (c) $a_1 = 20.0$. Overlaid are predictions from linear models $z = \hat{a}_0 + \hat{a}_1 x$ fitted on a 2 000-observation training set (not shown) using the SE loss (red lines), the $L_{\mathrm{NR}_2}$ loss (blue lines) and the $L_W$ loss (green lines).

We first detail the procedure for the dataset corresponding to $a_1 = 0.6$. We split the data into a training set of 2 000 observations and a test set of 2 000 observations. We then fit linear models of the form $z = a_0 + a_1 x$ to the training set using three different loss functions.

(i) Model #1: Fit using the SE loss (least squares regression).

(ii) Model #2: Fit using the $L_{\mathrm{NR}_2}$ loss.

(iii) Model #3: Fit using the $L_W$ loss.

The estimated parameters for model #1 are $\hat{a}_{01}, \hat{a}_{11}$ (given by eq. (3.21)), for model #2, they are $\hat{a}_{02}, \hat{a}_{12}$ (given by eq. (3.19)) and for model #3, they are $\hat{a}_{03}, \hat{a}_{13}$ (obtained by numerical minimization, as explicit estimators have not been derived; see Section 3.1.4). We then generate predictions for the test set using the equation $z = \hat{a}_0 + \hat{a}_1 x$. We repeat this entire procedure for the other two true values of $a_1$, while keeping the parameters of the probability distributions and the intercept fixed. Figure 4 presents the predictions from all models in the test set and Table 3 summarizes their numerical performance.

Table 3. Estimated parameters and test set performance for the three linear models from Section 4.2. For each value of the slope coefficient $a_1$, the table shows the parameters estimated on the training set and the subsequent performance metrics computed on the test set observations $\boldsymbol{y} = \boldsymbol{y}_{2\,001:4\,000}$. Predictions $\boldsymbol{z}$ are generated by each of the three fitted models.

| Parameter/ Metric | model #1 $a_1 = 0.6$ | model #2 $a_1 = 0.6$ | model #3 $a_1 = 0.6$ | model #1 $a_1 = 6.0$ | model #2 $a_1 = 6.0$ | model #3 $a_1 = 6.0$ | model #1 $a_1 = 2.1$ | model #2 $a_1 = 2.1$ | model #3 $a_1 = 2.1$ |
|---|---|---|---|---|---|---|---|---|---|
| $\hat{a}_0$ | 8.0655037 | -5.10243 | -16.74804 | 8.065504 | -2.011659 | -6.295647 | 8.065504 | 2.456986 | 1.750401 |
| $\hat{a}_1$ | 0.8935388 | 19.82837 | 12.47191 | 6.293539 | 20.783999 | 20.352048 | 20.293539 | 28.358309 | 28.347614 |
| $\mathrm{MSE}(\boldsymbol{z}, \boldsymbol{y})$ | 580.3332 | 996.0642 | 1002.823 | 580.3332 | 813.2273 | 818.57 | 580.3332 | 641.3484 | 641.8254 |
| $L_{\mathrm{NR}_2}(\boldsymbol{z}, \boldsymbol{y})$ | 0.9096833 | 0.4514271 | 0.4692202 | 0.5399942 | 0.332705 | 0.335796 | 0.1736741 | 0.1425356 | 0.1426502 |
| $L_W(\boldsymbol{z}, \boldsymbol{y})$ | 0.9657116 | 0.7132495 | 0.7077607 | 0.6732447 | 0.468934 | 0.466139 | 0.1963293 | 0.1620158 | 0.1619878 |
| $\bar{V}_{\mathrm{mean}}(\boldsymbol{z}, \boldsymbol{y})$ | -0.367136 | 0.4266737 | -16.5665 | -0.367136 | 0.371751 | -4.226231 | -0.367136 | 0.3898335 | -0.3245264 |

The predictions in Figure 4 and the parameter estimates in Table 3 show that the estimated models differ for lower values of $a_1$, but converge toward identical solutions as $a_1$ increases. For models #1 and #2, this behavior is explained by eqs. (3.19) and (3.21). As $a_1$ increases, the correlation $\rho(\boldsymbol{x}, \boldsymbol{y})$ tends toward 1, causing $\mathrm{sign}(\rho(\boldsymbol{x}, \boldsymbol{y})$ to become equal to $\rho(\boldsymbol{x}, \boldsymbol{y})$.

The model comparisons under each loss function are consistent with the theoretical framework established by [1], [24], as discussed in Section 2.1.5. Specifically, each model outperforms the others with respect to the loss function used for its training. Model #1 achieves the lowest MSE, model #2 the lowest empirical $L_{\mathrm{NR}_2}$ loss and model #3 the lowest empirical $L_W$ loss. As $a_1$ increases, the models become increasingly similar, and their performance metrics converge. The values of the $L_{\mathrm{NR}_2}$ and $L_W$ losses are very similar, which we attribute to the limited range [0,1] of these functions, as well as a potential equivalence for predictor-dependent variables with specific properties.

The numerical experiment also confirms that model #1 accurately identifies the conditional mean, as theoretically expected from Sections 2.1.4 and 2.1.5) (the value of $\bar{V}_{\mathrm{mean}}$ is nearly 0, consistently across all values of $a_1$). Notably, model #2 also exhibits this property, although this does not imply it can theoretically identify the conditional mean. In contrast, model #3 does not appear to identify the mean, at least for low values of $a_1$. As $a_1$ increases, its behavior becomes similar to that of #1 and #2.

### 4.3 Application to hydrologic modeling

To assess the practical relevance of our theoretical analysis in a hydrologic context, we applied the methods using the Génie Rural (GR) GR4J daily lumped hydrologic model [32]. The model was implemented via the `airGR` R package ([33], [34]). We modified the software to include the SE, $L_{\mathrm{NR}_2}$ and $L_W$ loss functions as options for model extremum estimation as outlined in see Section 2.2.

We applied the model to daily precipitation, potential evapotranspiration and streamflow data from 10 catchments in France. These catchments were selected from the `airGRdatasets` R package [35] based on the availability of complete records without missing values. Since these results are broadly applicable to various hydrologic models and catchments as well as machine learning regression models, we refrain from providing exhaustive details on the models and data herein. The software implementation is available as Supplementary Material.

The implementation was straightforward. We used the year 1999 as a warm-up period, the period from 2000-2008 for model calibration, and 2009-2018 for model testing (validation). For each catchment, we estimated three model instances, each using a different loss function:

(i) Model #1: Estimated using the SE loss (least squares regression).

(ii) Model #2: Estimated using the $L_{\mathrm{NR}_2}$ loss.

(iii) Model #3: Estimated using the $L_W$ loss.

Figure 5 presents examples of streamflow predictions from all three models for the year 2018 at catchment A273011002 (other years are omitted for visual clarity). The predictions from all models are highly similar. This visual similarity is confirmed by Figure 6, which plots observations against predictions for each model, showing a close alignment to the 1:1 line for all three.

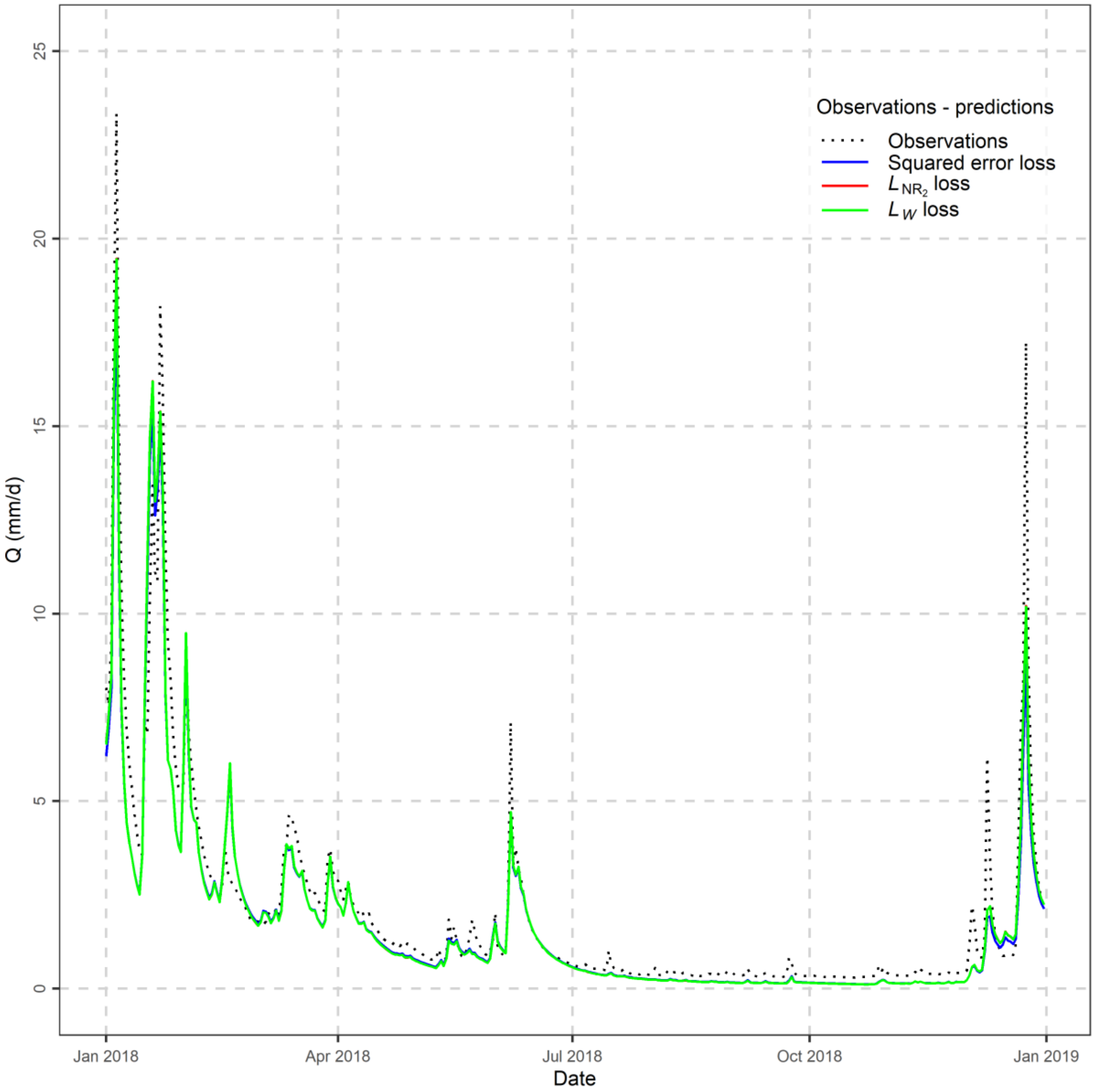

Figure 5. Predictions from the three models detailed in Section 4.3 for the year 2018 at catchment A273011002.

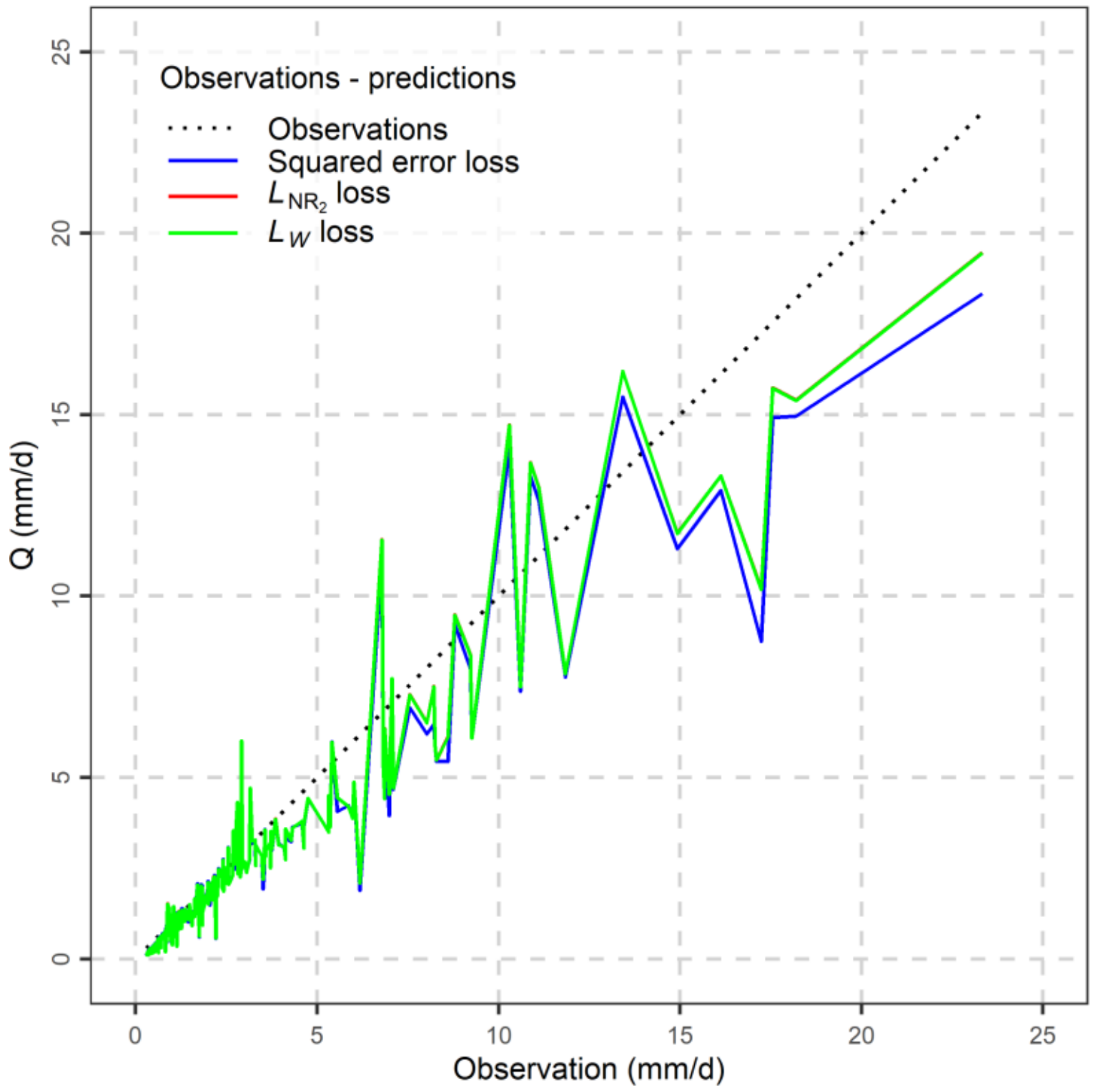

Figure 6. Observations plotted against predictions for the three models detailed in Section 4.3 for the year 2018 at catchment A273011002.

To quantify the differences between the models, we report their performance metrics for the calibration (estimation) period in Table 4 and for the validation (testing) period in Table 5. Presenting results for both periods is important due to the potential for statistical dissimilarity between the calibration and validation datasets, a factor not present in the controlled simulation setting of Section 4.2.

Table 4. Average losses (MSE, $L_{\mathrm{NR}_2}$, $L_W$) for all three models for each catchment in the calibration set. The best-performing model for each metric and catchment is highlighted in bold.

| Model | Model #1 | Model #2 | Model #3 | Model #1 | Model #2 | Model #3 | Model #1 | Model #2 | Model #3 |
|---|---|---|---|---|---|---|---|---|---|
| Basin/loss | MSE | MSE | MSE | $L_{\mathrm{NR}_2}$ | $L_{\mathrm{NR}_2}$ | $L_{\mathrm{NR}_2}$ | $L_W$ | $L_W$ | $L_W$ |
| A273011002 | **0.8004038** | 0.8213370 | 0.8206815 | 0.0412526 | **0.0402055** | 0.0402058 | 0.0428323 | 0.0417655 | **0.0417652** |
| A605102001 | **0.6815371** | 0.7118592 | 0.7116614 | 0.0574751 | **0.0549629** | 0.0549630 | 0.0604838 | 0.0578356 | **0.0578355** |
| B222001001 | **0.1452777** | 0.1473876 | 0.1472875 | 0.0240881 | **0.0237482** | 0.0237484 | 0.0246229 | 0.0242898 | **0.0242896** |
| F439000101 | **0.0156214** | 0.0158345 | 0.0158534 | 0.0220948 | **0.0217949** | 0.0217955 | 0.0225355 | 0.0222006 | **0.0222001** |
| H010002001 | **0.1543760** | 0.1565821 | 0.1565649 | 0.0280440 | **0.0276462** | **0.0276462** | 0.0287580 | 0.0283501 | 0.0283500 |
| H120101001 | **0.0987467** | 0.0999243 | 0.0998982 | 0.0225813 | **0.0223170** | 0.0223171 | 0.0230437 | 0.0227771 | **0.0227770** |
| H622101001 | **0.0719342** | 0.0725890 | 0.0725868 | 0.0151390 | **0.0150013** | **0.0150013** | 0.0153548 | **0.0152148** | **0.0152148** |
| J171171001 | **0.0999581** | 0.1008159 | 0.1008084 | 0.0159707 | **0.0158368** | **0.0158368** | 0.0162061 | **0.0160709** | **0.0160709** |
| J421191001 | **0.1908086** | 0.1919245 | 0.1919046 | 0.0109564 | **0.0108926** | **0.0108926** | 0.0110576 | **0.0109942** | **0.0109942** |
| K134181001 | **0.1093134** | 0.1104352 | 0.1103770 | 0.0167198 | **0.0165522** | 0.0165523 | 0.0169642 | 0.0168023 | **0.0168021** |

The results in the validation period (Table 5) were less uniform. For instance, Model #2 outperformed Model #1 with respect to MSE in 7 out of 10 catchments. This is expected, given the close similarity of the parameter estimates across all models and the potential for statistical dissimilarity between the calibration and validation periods.

Table 5. Average losses (MSE, $L_{\mathrm{NR}_2}$, $L_W$) for all three models for each catchment in the validation (test) set. The best-performing model for each metric and catchment is highlighted in bold.

| Model | Model #1 | Model #2 | Model #3 | Model #1 | Model #2 | Model #3 | Model #1 | Model #2 | Model #3 |
|---|---|---|---|---|---|---|---|---|---|
| Basin/loss | MSE | MSE | MSE | $L_{\mathrm{NR}_2}$ | $L_{\mathrm{NR}_2}$ | $L_{\mathrm{NR}_2}$ | $L_W$ | $L_W$ | $L_W$ |
| A273011002 | 0.7685125 | **0.7506903** | 0.7507642 | 0.0454044 | **0.0422782** | 0.0423143 | 0.0471956 | **0.0439308** | 0.0439685 |
| A605102001 | **0.4491356** | 0.4854598 | 0.4855521 | 0.0443461 | **0.0437923** | 0.0438100 | 0.0461854 | **0.0456384** | 0.0456576 |
| B222001001 | 0.1360638 | **0.1306809** | 0.1307524 | 0.0243684 | **0.0227379** | 0.0227663 | 0.0248823 | **0.0232137** | 0.0232427 |
| F439000101 | **0.0315423** | 0.0320811 | 0.0320874 | 0.0343641 | 0.0339732 | **0.0339406** | 0.0355261 | 0.0350892 | **0.0350533** |
| H010002001 | 0.1590732 | **0.1455477** | 0.1455789 | 0.0211302 | **0.0188218** | 0.0188279 | 0.0214761 | **0.0191156** | 0.0191217 |
| H120101001 | 0.2413662 | **0.2227167** | 0.2229445 | 0.0348958 | **0.0316126** | 0.0316505 | 0.0356209 | **0.0322373** | 0.0322764 |
| H622101001 | 0.1128062 | **0.1053143** | 0.1053167 | 0.0299121 | **0.0274271** | 0.0274283 | 0.0305397 | **0.0279809** | 0.0279821 |
| J171171001 | 0.1318598 | **0.1262418** | 0.1262628 | 0.0182125 | **0.0171466** | 0.0171507 | 0.0184918 | **0.0174056** | 0.0174098 |
| J421191001 | 0.2334413 | **0.2259492** | 0.2259939 | 0.0110676 | **0.0105920** | 0.0105951 | 0.0111783 | **0.0106949** | 0.0106981 |
| K134181001 | **0.0996014** | 0.1009891 | 0.1009341 | 0.0136642 | 0.0135830 | **0.0135825** | 0.0138255 | 0.0137501 | **0.0137495** |

During the calibration period (Table 4), each model performed best with respect to the specific loss function used for its training, as expected. However, the empirical loss values differed only minimally between models. This suggests that for this catchment dataset, the relationship between inputs and outputs is sufficiently strong that the different estimators converge to similar parameter sets, analogous to the high-correlation scenario illustrated in Figure 4c. Notably, in some cases the parameter estimates were so similar that models #2 and #3 achieved nearly identical performance under both the $L_{\mathrm{NR}_2}$ and $L_W$ losses.

Table 6 shows the empirical average identification function $\bar{V}_{\mathrm{mean}}$ for the mean, computed on the validation set. As expected from theory (Section 2.1.4), model #1, estimated with the SE loss, correctly identifies the mean (values near zero), indicating adequate absolute performance for mean prediction. Although not proven theoretically, Models #2 and #3 also produced values of $\bar{V}_{\mathrm{mean}}$ near zero. This is a consequence of their parameter estimates being very similar to those of model #1, as noted in the discussion of Table 4.

Table 6. Empirical average identification function $\bar{V}_{\mathrm{mean}}$ for all three models for each catchment in the validation set. The value closest to zero (indicating best identification of the mean) for each catchment is highlighted in bold.

| | Model | | |
|---|---|---|---|
| Basin | Model #1 | Model #2 | Model #3 |
| A273011002 | -0.1749492 | **-0.1325392** | -0.1332546 |
| A605102001 | -0.0438793 | **-0.0091700** | -0.0101047 |
| B222001001 | -0.0272470 | **-0.0061254** | -0.0065730 |
| F439000101 | -0.0570816 | -0.0526462 | **-0.0523539** |
| H010002001 | -0.0591676 | **-0.0388627** | -0.0389624 |
| H120101001 | -0.1592403 | **-0.1394463** | -0.1397120 |
| H622101001 | -0.1033602 | -0.0914236 | **-0.0913975** |
| J171171001 | -0.0452491 | **-0.0329591** | -0.0330311 |
| J421191001 | -0.1414669 | **-0.1233617** | -0.1234942 |
| K134181001 | **0.0775674** | 0.0939171 | 0.0936123 |

## 5. Discussion

At first glance, it may appear counterintuitive that $L_{\mathrm{NR}_2}$, when used for parameter estimation of a distribution, yields two distinct estimators, that differ from the mean. This contrasts with the empirical finding that, when used to fit a linear model, its predictions appear to equal the marginal mean of the dependent variable $\underline{y}$. This apparent contradiction may be explained by the fact that $L_{\mathrm{NR}_2}$ is defined only when $\boldsymbol{y}$ is non-constant or $\boldsymbol{z} \neq \boldsymbol{y}$. Consequently, the expectation $\mathbb{E}_F[L_{\mathrm{NR}_2}(z, \underline{y})]$ is defined only for distributions where $\boldsymbol{y}$ is non-constant. This restriction may introduce irregularities if one attempts its minimization as in eq. (2.4). This restriction may be circumvented when fitting a linear model, which allows for non-constant predictions $\boldsymbol{z}$ (for which $L_{\mathrm{NR}_2}$ is defined irrespectively of the distribution of $\underline{y}$). Similarly, $L_W$ does not identify a specific functional when fitting a probability distribution and its predictions do not necessarily have a mean equal to the marginal mean of $\underline{y}$.

In environmental sciences, it is sometimes assumed that predictions yielding a loss function value of 0 will perfectly predict the environmental process of interest, implying that a lower average loss invariably corresponds to a better prediction of the process itself. This concept must be understood precisely. Under the framework of expectation minimization of the loss function (eq. (2.4)), models predict specific functionals of the process rather than the process itself. For example, a linear model fitted with a quantile loss will predict conditional quantiles, not the process, even though the quantile loss equals 0 for a perfect prediction $\boldsymbol{z} = \boldsymbol{y}$ [36]. Applying this to our study, consider the analysis in Section 4.1. The mean climatology appears to better predict the process itself, yet it achieves higher (worse) $L_{\mathrm{NR}_2}$ and $L_W$ losses compared to predictions of the form $\mu(\boldsymbol{y}) \pm \sigma(\boldsymbol{y})$. All losses equal 0 only for a perfect prediction $\boldsymbol{z} = \boldsymbol{y}$. This result is less paradoxical when considering that models predict functionals of the predictive distribution rather than the process itself. That principle already applied in hydrologic science and engineering to predict quantiles ([37], [38], [39]) and expectiles (an extension of the mean functional [40]) using hydrologic models [41].

In environmental sciences and engineering, it is common practice to assess predictions using multiple metrics, and debates continue over which metrics are superior [8]. This perspective has driven the introduction of numerous new metrics. This practice should be contextualized within statistical theory, which advocates for the a priori specification of the loss function used for model assessment ([1], [24]), as model rankings are critically dependent on this choice. Our work underscores the importance of this approach, while establishing the theoretical properties of the $L_{\mathrm{NR}_2}$ and $L_W$ loss functions. In this context, we do not argue for the universal superiority of one loss function. Each loss function has distinct theoretical merits and, more importantly, each is intrinsically linked to the specific functional we aim to predict. The SE loss is simpler to use and highly interpretable, as it elicits the mean. The $L_{\mathrm{NR}_2}$ loss yields specific estimates when used as an extremum estimator on a sample from a probability distribution, though it lacks consistency due to non-uniqueness. However, when applied to a linear model, it provides unique parameter estimates, and our empirical evaluation suggests that the expectation of its predictions equals the mean of the marginal distribution of the dependent variable. A significant advantage of $L_{\mathrm{NR}_2}$ is its translation and scale invariance, coupled with its bounded range of [0,1]. In environmental science and engineering, such boundedness is often perceived to enhance the interpretability of prediction assessments. $L_W$ also possesses translation and scale invariance and is bounded in [0,1]. However, it likely lacks the property of predicting the marginal mean of the dependent variable, when used to fit a linear model and does not yield explicit parameter estimates.

It would be interesting to examine the consistency of $L_{\mathrm{NR}_2}$ and $L_W$ for multidimensional functionals (our analysis was restricted to one-dimensional functionals), a topic discussed by [1]. For instance, the general form of a consistent loss function for the component-wise expectation is a Bregman loss function ([12], [16], [42],)

$$L(\boldsymbol{z}, \boldsymbol{y}) := \varphi(\boldsymbol{y}) - \varphi(\boldsymbol{z}) - \langle \nabla\varphi(\boldsymbol{z}), \boldsymbol{y} - \boldsymbol{z} \rangle \tag{5.1}$$

where $\varphi: \mathbb{R}^d \to \mathbb{R}$ is convex with gradient $\nabla\varphi: \mathbb{R}^d \to \mathbb{R}^d$. The SE loss is a special case of eq. (5.1). In this regard, a challenge for both $L_{\mathrm{NR}_2}$ and $L_W$ is that they must be examined within the class of distributions for which $\boldsymbol{y}$ is not constant.

Preliminary theoretical and empirical work by the authors on $L_{\mathrm{NR}_p}$ has not identified specific properties related to existing functionals. Determining what functional, if any, models fitted with $L_{\mathrm{NR}_p}$ identify remains an open problem. Such functionals, if they exist, are termed hydrograph functionals by [43] due to their significance in hydrologic science and engineering.

## 6. Conclusions

In this study, we examined the theoretical properties of the index of agreement loss function, $L_W$ (the negatively oriented index of agreement). Our analysis established that $L_W$is translation and scale invariant, bounded within [0,1], and, when employed as an extremum estimator for a sample from a probability distribution, yields the parameter estimates $\mathbb{E}_F[\underline{y}] \pm \mathbb{V}_F^{1/2}[\underline{y}]$.

We proposed $L_{\mathrm{NR}_2}$ as a theoretically consistent improvement over the original index of agreement. Unlike $L_W$, whose denominator relies on component-wise vector operations, $L_{\mathrm{NR}_2}$ is constructed using Euclidean distances in both its numerator and denominator. This new formulation retains the appealing properties of translation and scale invariance and a [0,1] bound and it produces the same $\mathbb{E}_F[\underline{y}] \pm \mathbb{V}_F^{1/2}[\underline{y}]$ estimates when fitting a distribution. A significant strength of $L_{\mathrm{NR}_2}$ is that it provides closed-form solutions when used as an extremum estimator to fit a linear model.

Empirical analysis demonstrated that predictions of linear models fitted with $L_{\mathrm{NR}_2}$ have mean equal to the mean of the dependent variable, a property shared with the squared error loss function (though the latter identifies the conditional mean). This represents a notable benefit over $L_W$, for which we found no evidence of a similar identification property.

When applied to calibrate hydrologic models, both loss functions resulted in parameter sets that produced remarkably similar performance across all empirical average loss metrics, including MSE. This behavior is analogous to the high-correlation scenario in linear regression, where the models converge as the correlation $\rho(\boldsymbol{x}, \boldsymbol{y})$ approaches 1.

Finally, while variants of the $L_W$ loss based on norms other than the Euclidean norm have been proposed, we introduced a corresponding generalization, $L_{\mathrm{NR}_p}$, based on the principles underlying $L_{\mathrm{NR}_2}$. The theoretical properties of this generalized function remain an open area of research. Preliminary results did not indicate that models trained with the $L_{\mathrm{NR}_1}$ loss, for instance, can identify the median or mean functional.

## Appendix A Vector notation

This appendix defines the vector operations, norms, and statistical functions used throughout the manuscript. All vectors are column vectors in $\mathbb{R}^n$.

## Basic vector definitions

Vector notation:

$$\boldsymbol{x} = (x_1, \dots, x_n)^{\mathrm{T}} \tag{A.1}$$

where the superscript $^{\mathrm{T}}$ denotes the transpose.

Zero vector (all elements zero):

$$\mathbf{0} = (0, \dots, 0)^{\mathrm{T}} \tag{A.2}$$

All-ones vector (all elements unity):

$$\mathbf{1} = (1, \dots, 1)^{\mathrm{T}} \tag{A.3}$$

## Element-wise operations

Absolute value vector:

$$|\boldsymbol{x}| := (|x_1|, \dots, |x_n|)^{\mathrm{T}} \tag{A.4}$$

Element-wise comparison (holds for all $i \in \{1, \dots, n\}$):

$$\boldsymbol{x} > \boldsymbol{y} \Leftrightarrow x_i > y_i \text{ (similarly for } \geq, =, \leq, <) \tag{A.5}$$

Element-wise inequality:

$$\boldsymbol{x} \neq \boldsymbol{y} \Leftrightarrow \exists i \in \{1, \dots, n\}: x_i \neq y_i \tag{A.6}$$

## Inner products and norms

Euclidean inner product:

$$\langle \boldsymbol{x}, \boldsymbol{y} \rangle := \boldsymbol{x}^{\mathrm{T}} \boldsymbol{y} = \sum_{i=1}^{n} x_i y_i = \langle \boldsymbol{y}, \boldsymbol{x} \rangle \tag{A.7}$$

$L_p$-norms ($p \geq 1$):

$$\|\boldsymbol{x}\|_1 = \sum_{i=1}^{n} |x_i| \text{ (Manhattan norm)} \tag{A.8}$$

$$\|\boldsymbol{x}\|_2 = \left(\sum_{i=1}^{n} x_i^2\right)^{1/2} \text{ (Euclidean norm)} \tag{A.9}$$

$$\|\boldsymbol{x}\|_p := \left(\sum_{i=1}^{n} |x_i|^p\right)^{1/p} \text{ (}L_p\text{-norm)} \tag{A.10}$$

$$\|\boldsymbol{x}\|_\infty = \max\{|x_1|, \dots, |x_n|\} \text{ (Chebyshev norm)} \tag{A.11}$$

Squared Euclidean norm:

$$\|\boldsymbol{x}\|_2^2 = \langle \boldsymbol{x}, \boldsymbol{x} \rangle = \boldsymbol{x}^{\mathrm{T}} \boldsymbol{x} = \sum_{i=1}^{n} x_i^2 \tag{A.12}$$

Scaling invariance:

$$\|a\boldsymbol{x}\|_2^2 = a^2 \|\boldsymbol{x}\|_2^2 \tag{A.13}$$

Norm equivalences:

$$\||\boldsymbol{x}|\|_2 = \|\boldsymbol{x}\|_2 \tag{A.14}$$

$$\langle |\boldsymbol{x}|, |\boldsymbol{x}| \rangle = \langle \boldsymbol{x}, \boldsymbol{x} \rangle = \|\boldsymbol{x}\|_2^2 \tag{A.15}$$

**Sign and indicator functions**

Scalar sign:

$$\operatorname{sign}(x) := \begin{cases} 1, x > 0 \\ -1, x < 0 \end{cases} \tag{A.16}$$

Vector extension:

$$\operatorname{sign}(\boldsymbol{x}) := (\operatorname{sign}(x_1), \dots, \operatorname{sign}(x_n))^{\mathrm{T}} \tag{A.17}$$

Indicator function:

$$\mathbb{1}_A\{x\} := \begin{cases} 1, x \in A \\ 0, \text{otherwise} \end{cases} \tag{A.18}$$

**Statistical functions for vectors**

Sample mean:

$$\mu(\boldsymbol{x}) := (1/n)\mathbf{1}^{\mathrm{T}}\boldsymbol{x} = (1/n)\textstyle\sum_{i=1}^{n} x_i = (1/n)\langle \boldsymbol{x}, \mathbf{1} \rangle \tag{A.19}$$

Mean absolute deviation:

$$\operatorname{mad}(\boldsymbol{x}) := \mu(|\mathbf{1}\mu(\boldsymbol{x}) - \boldsymbol{x}|) \tag{A.20}$$

Sample standard deviation:

$$\sigma(\boldsymbol{x}) := \sqrt{(\|\boldsymbol{x}\|_2^2 - n\mu^2(\boldsymbol{x}))}/\sqrt{n} \tag{A.21}$$

Sample variance:

$$\sigma^2(\boldsymbol{x}) = (\|\boldsymbol{x}\|_2^2 - n\mu^2(\boldsymbol{x}))/n \tag{A.22}$$

**Centered vectors and key identities:**

Centered vector (mean-zero):

$$\boldsymbol{x}_c := \boldsymbol{x} - \mathbf{1}\mu(\boldsymbol{x}) \tag{A.23}$$

Pythagorean decomposition:

$$\|\boldsymbol{x}\|_2^2 = \|\mathbf{1}\mu(\boldsymbol{x})\|_2^2 + \|\boldsymbol{x}_{\mathrm{c}}\|_2^2 \tag{A.24}$$

Sum of squared deviations:

$$\|\boldsymbol{x}_{\mathrm{c}}\|_2^2 = \textstyle\sum_{i=1}^{n} (x_i - \mu(\boldsymbol{x}))^2 \tag{A.25}$$

Norm expansion:

$$\|\boldsymbol{x} + \boldsymbol{y}\|_2^2 = \|\boldsymbol{x}\|_2^2 + \|\boldsymbol{y}\|_2^2 + 2\langle \boldsymbol{x}, \boldsymbol{y} \rangle \tag{A.26}$$

Orthogonality to unity:

$$\langle \boldsymbol{x}_{\mathrm{c}}, \mathbf{1} \rangle = 0 \tag{A.27}$$

Pearson correlation:

$$\rho(\boldsymbol{x},\boldsymbol{y}) := \langle \boldsymbol{x}_{\mathrm{c}}, \boldsymbol{y}_{\mathrm{c}} \rangle / (\|\boldsymbol{x}_{\mathrm{c}}\|_2 \|\boldsymbol{y}_{\mathrm{c}}\|_2) \tag{A.28}$$

**<u>Inequalities:</u>**

Cauchy-Schwarz inequality ([44], p.33):

$$|\langle \boldsymbol{x},\boldsymbol{y} \rangle| \leq \|\boldsymbol{x}\|_2 \|\boldsymbol{y}\|_2 \tag{A.29}$$

with equality ([44], p.34):

$$|\langle \boldsymbol{x},\boldsymbol{y} \rangle| = \|\boldsymbol{x}\|_2 \|\boldsymbol{y}\|_2, \text{ iff } \boldsymbol{x} = \boldsymbol{0}, \text{ or } \boldsymbol{y} = \boldsymbol{0}, \text{ or } \boldsymbol{x} = a\boldsymbol{y} \tag{A.30}$$

Triangle inequality ([44], p.35):

$$\|\boldsymbol{x}+\boldsymbol{y}\|_2 \leq \|\boldsymbol{x}\|_2 + \|\boldsymbol{y}\|_2 \tag{A.31}$$

with equality

$$\|\boldsymbol{x}+\boldsymbol{y}\|_2 = \|\boldsymbol{x}\|_2 + \|\boldsymbol{y}\|_2, \text{ iff } \boldsymbol{x} = \boldsymbol{0}, \text{ or } \boldsymbol{y} = \boldsymbol{0}, \text{ or } \boldsymbol{x} = a\boldsymbol{y}, a > 0 \tag{A.32}$$

Norm bounds:

$$\|\boldsymbol{x}\|_2 \leq \|\boldsymbol{x}\|_1 \leq \sqrt{n}\|\boldsymbol{x}\|_2 \tag{A.33}$$

**<u>Variance identity</u>**

$$\|\boldsymbol{1}\mu(\boldsymbol{x}) - \boldsymbol{x}\|_2^2 = \langle \boldsymbol{1}\mu(\boldsymbol{x}) - \boldsymbol{x}, \boldsymbol{1}\mu(\boldsymbol{x}) - \boldsymbol{x} \rangle \Rightarrow \tag{A.34}$$

$$\|\boldsymbol{1}\mu(\boldsymbol{x}) - \boldsymbol{x}\|_2^2 = \|\boldsymbol{1}\|_2^2\, \mu^2(\boldsymbol{x}) + \|\boldsymbol{x}\|_2^2 - 2\mu(\boldsymbol{x})\langle \boldsymbol{1}, \boldsymbol{x} \rangle \Rightarrow \tag{A.35}$$

$$\|\boldsymbol{1}\mu(\boldsymbol{x}) - \boldsymbol{x}\|_2^2 = n\mu^2(\boldsymbol{x}) + \|\boldsymbol{x}\|_2^2 - 2n\mu^2(\boldsymbol{x}) \Rightarrow \tag{A.36}$$

$$\|\boldsymbol{1}\mu(\boldsymbol{x}) - \boldsymbol{x}\|_2^2 = \|\boldsymbol{x}\|_2^2 - n\mu^2(\boldsymbol{x}) \Rightarrow \tag{A.37}$$

$$\|\boldsymbol{1}\mu(\boldsymbol{x}) - \boldsymbol{x}\|_2^2 = n\sigma^2(\boldsymbol{x}) \tag{A.38}$$

## Appendix B Loss and identification functions

This appendix defines the established loss and identification functions referenced throughout this paper. Let $\boldsymbol{z} = (z_1, \dots, z_n)^{\mathrm{T}}$ be a vector of predictions and $\boldsymbol{y} = (y_1, \dots, y_n)^{\mathrm{T}}$ be a vector of corresponding observed values.

The absolute error loss function is defined as:

$$L_{\mathrm{AE}}(z,y) := |z-y| \tag{B.1}$$

The squared error loss function is defined as:

$$L_{\mathrm{SE}}(z,y) := (z-y)^2 \tag{B.2}$$

The mean absolute error (MAE) is defined as:

$$\mathrm{MAE}(\boldsymbol{z},\boldsymbol{y}) := (1/n)\sum_{i=1}^{n} L_{\mathrm{AE}}(z_i, y_i) \tag{B.3}$$

The mean squared error (MSE) is defined as:

$$\mathrm{MSE}(\boldsymbol{z},\boldsymbol{y}) := (1/n)\sum_{i=1}^{n} L_{\mathrm{SE}}(z_i, y_i) \tag{B.4}$$

The Nash-Sutcliffe efficiency (NSE) [17] is defined as:

$$\mathrm{NSE}(\boldsymbol{z},\boldsymbol{y}) := 1 - \mathrm{MSE}(\boldsymbol{z},\boldsymbol{y})/\mathrm{MSE}(\mathbf{1}\mu(\boldsymbol{y}),\boldsymbol{y}) \tag{B.5}$$

A strict identification function for the median [1] is defined as:

$$V_{\mathrm{median}}(z,y) := \mathbb{1}_{[0,\infty)}\{z-y\} - 1/2 \tag{B.6}$$

where $\mathbb{1}_A\{x\}$ is the indicator function defined in eq. (A.18).

A strict identification function for the mean [1] is defined as:

$$V_{\mathrm{mean}}(z,y) := z - y \tag{B.7}$$

The empirical average identification function $\bar{V}_{\mathrm{median}}$ is defined as:

$$\bar{V}_{\mathrm{median}}(\boldsymbol{z},\boldsymbol{y}) = (1/n)\sum_{i=1}^{n} V_{\mathrm{median}}(z_i, y_i) \tag{B.8}$$

The empirical average identification function $\bar{V}_{\mathrm{mean}}$ is defined as:

$$\bar{V}_{\mathrm{mean}}(\boldsymbol{z},\boldsymbol{y}) = (1/n)\sum_{i=1}^{n} V_{\mathrm{mean}}(z_i, y_i) \tag{B.9}$$

## Appendix C Probability distributions

This appendix defines the probability density functions (PDFs) for the probability distributions referenced in the main text.

The PDF of a Gaussian (normal) random variable $\underline{y} \sim N(\mu,\sigma)$ is defined as:

$$f_N(y;\mu,\sigma) := \frac{1}{\sigma\sqrt{2\pi}}\exp\left(-\frac{(y-\mu)^2}{2\sigma^2}\right), y \in \mathbb{R}, \mu \in \mathbb{R}, \sigma > 0 \tag{C.1}$$

The PDF of a Gamma distributed random variable $\underline{y} \sim \mathrm{Gamma}(\theta, a)$ is defined as:

$$f_{\mathrm{Gamma}}(y;\theta,a) := \frac{1}{\theta^a\Gamma(a)} y^{a-1}\exp(-y/\theta), y \geq 0, \theta > 0, a > 0 \tag{C.2}$$

The PDF of a log-normally distributed random variable $\underline{y} \sim \mathrm{Lognormal}(\mu,\sigma)$ is defined as:

$$f_{\mathrm{Lognormal}}(y;\mu,\sigma) := \frac{1}{y\sigma\sqrt{2\pi}}\exp\left(-\frac{(\log(y)-\mu)^2}{2\sigma^2}\right), y > 0, \mu \in \mathbb{R}, \sigma > 0 \tag{C.3}$$

## Appendix D Proofs

**Proof D.1** (Bound of $L_W$, adapted from [45]): The bound $0 \leq L_W(\boldsymbol{z},\boldsymbol{y}) \leq 1$ is established as follows:

(i) Lower bound

The lower bound follows immediately, since both numerator and denominator are squared Euclidean norms, which are non-negative.

(ii) Upper bound

Starting from the element-wise triangle inequality for vectors (eq. (A.5)):

$$|\boldsymbol{z}-\boldsymbol{y}| = |\boldsymbol{z}-\mathbf{1}\mu(\boldsymbol{y})+\mathbf{1}\mu(\boldsymbol{y})-\boldsymbol{y}| \leq |\boldsymbol{z}-\mathbf{1}\mu(\boldsymbol{y})|+|\mathbf{1}\mu(\boldsymbol{y})-\boldsymbol{y}|, \quad \text{(D.1)}$$

where the inequality holds element-wise. Applying the Euclidian norm (which preserves order for non-negative vectors), squaring and by leveraging eq. (A.14) yields

$$\|\boldsymbol{z}-\boldsymbol{y}\|_2^2 \leq \|(|\boldsymbol{z}-\mathbf{1}\mu(\boldsymbol{y})|+|\mathbf{1}\mu(\boldsymbol{y})-\boldsymbol{y}|)\|_2^2 \quad \text{(D.2)}$$

Thus $L_W(\boldsymbol{z},\boldsymbol{y}) \leq 1$.■

**Proof D.2** (Translation invariance of $L_W$): Substituting translation vectors into eq. (2.13):

$$L_W(\boldsymbol{z}+c\mathbf{1},\boldsymbol{y}+c\mathbf{1}) = \|(\boldsymbol{z}+c\mathbf{1})-(\boldsymbol{y}+c\mathbf{1})\|_2^2/\|(|(\boldsymbol{z}+c\mathbf{1})-\mathbf{1}\mu(\boldsymbol{y}+c\mathbf{1})|+|\mathbf{1}\mu(\boldsymbol{y}+c\mathbf{1})-(\boldsymbol{y}+c\mathbf{1})|)\|_2^2 \Rightarrow \quad \text{(D.3)}$$

$$L_W(\boldsymbol{z}+c\mathbf{1},\boldsymbol{y}+c\mathbf{1}) = \|\boldsymbol{z}-\boldsymbol{y}\|_2^2/\|(|\boldsymbol{z}-\mathbf{1}\mu(\boldsymbol{y})|+|\mathbf{1}\mu(\boldsymbol{y})-\boldsymbol{y}|)\|_2^2 \Rightarrow \quad \text{(D.4)}$$

$$L_W(\boldsymbol{z}+c\mathbf{1},\boldsymbol{y}+c\mathbf{1}) = L_W(\boldsymbol{z},\boldsymbol{y}), c \in \mathbb{R} \quad \text{(D.5)}$$

Thus $L_W$ is translation invariant.■

**Proof D.3** ($L_W$ for extremum estimation): We show that $\underset{\theta\in\Theta}{\operatorname{argmin}}\, L_W(\theta\mathbf{1},\boldsymbol{y}) = \mu(\boldsymbol{y}) \pm \sigma(\boldsymbol{y})$.

For $\boldsymbol{z}=\theta\mathbf{1}$ and $\boldsymbol{y}$ not constant, the nominator of $L_W$ is:

$$\|\boldsymbol{z}-\boldsymbol{y}\|_2^2 = \|\theta\mathbf{1}-\boldsymbol{y}\|_2^2 \quad \text{(D.6)}$$

By expanding the norm (eq. (A.26)) it follows that:

$$\|\boldsymbol{z}-\boldsymbol{y}\|_2^2 = \|\theta\mathbf{1}\|_2^2+\|\boldsymbol{y}\|_2^2-2\langle\theta\mathbf{1},\boldsymbol{y}\rangle \Rightarrow \quad \text{(D.7)}$$

$$\|\boldsymbol{z}-\boldsymbol{y}\|_2^2 = n\theta^2+\|\boldsymbol{y}\|_2^2-2n\theta\mu(\boldsymbol{y}) \Rightarrow \quad \text{(D.8)}$$

$$\|\boldsymbol{z}-\boldsymbol{y}\|_2^2 = n(\theta-\mu(\boldsymbol{y}))^2+\|\boldsymbol{y}\|_2^2-n\mu^2(\boldsymbol{y}) \quad \text{(D.9)}$$

By substituting with the sample variance (eq. (A.22)) it follows that:

$$\|\boldsymbol{z}-\boldsymbol{y}\|_2^2 = n(\theta-\mu(\boldsymbol{y}))^2+n\sigma^2(\boldsymbol{y}) \quad \text{(D.10)}$$

The denominator expands to:

$$\|(|\boldsymbol{z}-\mathbf{1}\mu(\boldsymbol{y})|+|\mathbf{1}\mu(\boldsymbol{y})-\boldsymbol{y}|)\|_2^2 = \||\boldsymbol{z}-\mathbf{1}\mu(\boldsymbol{y})|\|_2^2+\||\mathbf{1}\mu(\boldsymbol{y})-\boldsymbol{y}|\|_2^2+2\langle|\boldsymbol{z}-\mathbf{1}\mu(\boldsymbol{y})|,|\mathbf{1}\mu(\boldsymbol{y})-\boldsymbol{y}|\rangle \Rightarrow \quad \text{(D.11)}$$

$$\|(|\boldsymbol{z}-\mathbf{1}\mu(\boldsymbol{y})|+|\mathbf{1}\mu(\boldsymbol{y})-\boldsymbol{y}|)\|_2^2 = \||\theta\mathbf{1}-\mathbf{1}\mu(\boldsymbol{y})|\|_2^2+\||\mathbf{1}\mu(\boldsymbol{y})-\boldsymbol{y}|\|_2^2+2\langle|\theta\mathbf{1}-\mathbf{1}\mu(\boldsymbol{y})|,|\mathbf{1}\mu(\boldsymbol{y})-\boldsymbol{y}|\rangle \quad \text{(D.12)}$$

The first term in eq. (D.12) is

$$|||\theta\mathbf{1}-\mathbf{1}\mu(\boldsymbol{y})|||_2^2=(\theta-\mu(\boldsymbol{y}))^2\|\mathbf{1}\|_2^2\Rightarrow \tag{D.13}$$

$$|||\theta\mathbf{1}-\mathbf{1}\mu(\boldsymbol{y})|||_2^2=n(\theta-\mu(\boldsymbol{y}))^2 \tag{D.14}$$

From eq. (A.38), the second term in eq. (D.12) is

$$|||\mathbf{1}\mu(\boldsymbol{y})-\boldsymbol{y}|||_2^2=n\sigma^2(\boldsymbol{y}) \tag{D.15}$$

The last term in eq. (D.12) is

$$\langle|\theta\mathbf{1}-\mathbf{1}\mu(\boldsymbol{y})|,|\mathbf{1}\mu(\boldsymbol{y})-\boldsymbol{y}|\rangle=|\theta-\mu(\boldsymbol{y})|\langle|\mathbf{1}|,|\mathbf{1}\mu(\boldsymbol{y})-\boldsymbol{y}|\rangle\Rightarrow \tag{D.16}$$

$$\langle|\theta\mathbf{1}-\mathbf{1}\mu(\boldsymbol{y})|,|\mathbf{1}\mu(\boldsymbol{y})-\boldsymbol{y}|\rangle=|\theta-\mu(\boldsymbol{y})|\langle\mathbf{1},|\mathbf{1}\mu(\boldsymbol{y})-\boldsymbol{y}|\rangle\Rightarrow \tag{D.17}$$

$$\langle|\theta\mathbf{1}-\mathbf{1}\mu(\boldsymbol{y})|,|\mathbf{1}\mu(\boldsymbol{y})-\boldsymbol{y}|\rangle=|\theta-\mu(\boldsymbol{y})|\|\mathbf{1}\mu(\boldsymbol{y})-\boldsymbol{y}\|_1\Rightarrow \tag{D.18}$$

$$\langle|\theta\mathbf{1}-\mathbf{1}\mu(\boldsymbol{y})|,|\mathbf{1}\mu(\boldsymbol{y})-\boldsymbol{y}|\rangle=n|\theta-\mu(\boldsymbol{y})|\mathrm{mad}(\boldsymbol{y}) \tag{D.19}$$

where $\mathrm{mad}(\boldsymbol{y})$ is the sample mean absolute deviation of $\boldsymbol{y}$ defined in (A.20). Then $L_W(\theta\mathbf{1},\boldsymbol{y})$ equals:

$$L_W(\theta\mathbf{1},\boldsymbol{y})=(n(\theta-\mu(\boldsymbol{y}))^2+n\sigma^2(\boldsymbol{y}))/(n(\theta-\mu(\boldsymbol{y}))^2+n\sigma^2(\boldsymbol{y})+2n|\theta-\mu(\boldsymbol{y})|\mathrm{mad}(\boldsymbol{y}))\Rightarrow \tag{D.20}$$

$$L_W(\theta\mathbf{1},\boldsymbol{y})=((\theta-\mu(\boldsymbol{y}))^2+\sigma^2(\boldsymbol{y}))/((\theta-\mu(\boldsymbol{y}))^2+\sigma^2(\boldsymbol{y})+2|\theta-\mu(\boldsymbol{y})|\mathrm{mad}(\boldsymbol{y})) \tag{D.21}$$

Let

$$u_1=\theta-\mu(\boldsymbol{y}) \tag{D.22}$$

$$u_2=\sigma(\boldsymbol{y}) \tag{D.23}$$

$$u_3=\mathrm{mad}(\boldsymbol{y}) \tag{D.24}$$

After substituting eqs. (D.22), (D.23) and (D.24) in (D.21) we have:

$$L_W(\theta\mathbf{1},\boldsymbol{y})=(u_1^2+u_2^2)/(u_1^2+u_2^2+2|u_1|u_3) \tag{D.25}$$

The partial derivative $\frac{\partial L_W(\theta\mathbf{1},\boldsymbol{y})}{\partial\theta}$ is defined for $\theta\neq\mu(\boldsymbol{y})$. For $\theta>\mu(\boldsymbol{y})$:

$$\frac{\partial L_W(\theta\mathbf{1},\boldsymbol{y})}{\partial\theta}=(2u_1(u_1^2+u_2^2+2u_1u_3)-2(u_1+u_3)(u_1^2+u_2^2))/(u_1^2+u_2^2+2u_1u_3)^2\Rightarrow \tag{D.26}$$

$$\frac{\partial L_W(\theta\mathbf{1},\boldsymbol{y})}{\partial\theta}=2u_3(u_1^2-u_2^2)/(u_1^2+u_2^2+2u_1u_3)^2 \tag{D.27}$$

For $\theta<\mu(\boldsymbol{y})$:

$$\frac{\partial L_W(\theta\mathbf{1},\boldsymbol{y})}{\partial\theta}=(2u_1(u_1^2+u_2^2-2u_1u_3)-2(u_1-u_3)(u_1^2+u_2^2))/(u_1^2+u_2^2-2u_1u_3)^2\Rightarrow \tag{D.28}$$

$$\frac{\partial L_W(\theta\mathbf{1},\boldsymbol{y})}{\partial\theta} = -2u_3(u_1^2 - u_2^2)/(u_1^2 + u_2^2 - 2u_1u_3)^2 \tag{D.29}$$

Combining the two partial derivatives, we have

$$\frac{\partial L_W(\theta\mathbf{1},\boldsymbol{y})}{\partial\theta} = (2u_3(u_1^2 - u_2^2)\mathrm{sign}(u_1))/(u_1^2 + u_2^2 + 2|u_1|u_3)^2 \tag{D.30}$$

which is a function of $\theta$ only through $u_1$. The sign function is defined in eq. (A.16). The partial derivative $\frac{\partial L_W(\theta\mathbf{1},\boldsymbol{y})}{\partial\theta}$ is not defined at $\theta = \mu(\boldsymbol{y})$ because (see Figure 1)

$$\lim_{\theta\to\mu(\boldsymbol{y})^+} \frac{\partial L_W(\theta\mathbf{1},\boldsymbol{y})}{\partial\theta} = -2u_3/u_2^2 \tag{D.31}$$

and

$$\lim_{\theta\to\mu(\boldsymbol{y})^-} \frac{\partial L_W(\theta\mathbf{1},\boldsymbol{y})}{\partial\theta} = 2u_3/u_2^2 \tag{D.32}$$

The second order partial derivative $\frac{\partial^2 L_W(\theta\mathbf{1},\boldsymbol{y})}{\partial\theta^2}$ is defined for $\theta \neq \mu(\boldsymbol{y})$.

We set

$$u_4 = u_1^2 + u_2^2 + 2u_1u_3 \tag{D.33}$$

$$u_5 = u_1^2 + u_2^2 - 2u_1u_3 \tag{D.34}$$

For $\theta > \mu(\boldsymbol{y})$:

$$\frac{\partial^2 L_W(\theta\mathbf{1},\boldsymbol{y})}{\partial\theta^2} = (4u_1u_3u_4^2 - 8u_3(u_1^2 - u_2^2)u_4(u_1 + u_3))/u_4^4 \Rightarrow \tag{D.35}$$

$$\frac{\partial^2 L_W(\theta\mathbf{1},\boldsymbol{y})}{\partial\theta^2} = (4u_3(u_1u_4 - 2(u_1^2 - u_2^2)(u_1 + u_3)))/u_4^3 \Rightarrow \tag{D.36}$$

$$\frac{\partial^2 L_W(\theta\mathbf{1},\boldsymbol{y})}{\partial\theta^2} = (4u_3(u_1^3 + u_1u_2^2 + 2u_1^2u_3 - 2(u_1^3 + u_1^2u_3 - u_1u_2^2 - u_3u_2^2)))/u_4^3 \Rightarrow \tag{D.37}$$

$$\frac{\partial^2 L_W(\theta\mathbf{1},\boldsymbol{y})}{\partial\theta^2} = (4u_3(-u_1^3 + 3u_1u_2^2 + 2u_3u_2^2))/u_4^3 \Rightarrow \tag{D.38}$$

$$\frac{\partial^2 L_W(\theta\mathbf{1},\boldsymbol{y})}{\partial\theta^2} = (4u_3(2u_3u_2^2 + u_1(3u_2^2 - u_1^2)))/u_4^3 \tag{D.39}$$

For $\theta < \mu(\boldsymbol{y})$:

$$\frac{\partial^2 L_W(\theta\mathbf{1},\boldsymbol{y})}{\partial\theta^2} = (-4u_1u_3u_5^2 + 8u_3(u_1^2 - u_2^2)u_5(u_1 - u_3))/u_5^4 \Rightarrow \tag{D.40}$$

$$\frac{\partial^2 L_W(\theta\mathbf{1},\boldsymbol{y})}{\partial\theta^2} = (4u_3(-u_1u_5 + 2(u_1^2 - u_2^2)(u_1 - u_3)))/u_5^3 \Rightarrow \tag{D.41}$$

$$\frac{\partial^2 L_W(\theta\mathbf{1},\boldsymbol{y})}{\partial\theta^2} = (4u_3(-u_1^3 - u_1u_2^2 + 2u_1^2u_3 + 2(u_1^3 - u_1^2u_3 - u_1u_2^2 + u_3u_2^2)))/u_5^3 \Rightarrow \tag{D.42}$$

$$\frac{\partial^2 L_W(\theta\mathbf{1},\mathbf{y})}{\partial\theta^2} = (4u_3(u_1^3 - 3u_1u_2^2 + 2u_3u_2^2))/u_5^3 \Rightarrow \quad \text{(D.43)}$$

$$\frac{\partial^2 L_W(\theta\mathbf{1},\mathbf{y})}{\partial\theta^2} = (4u_3(2u_3u_2^2 - u_1(3u_2^2 - u_1^2)))/u_5^3 \quad \text{(D.44)}$$

Combining the two partial derivatives, we have

$$\frac{\partial^2 L_W(\theta\mathbf{1},\mathbf{y})}{\partial\theta^2} = (4u_3(2u_3u_2^2 + u_1(3u_2^2 - u_1^2)\text{sign}(u_1)))/(u_1^2 + u_2^2 + 2|u_1|u_3)^3 \quad \text{(D.45)}$$

which is a function of $\theta$ only through $u_1$. The partial derivative $\frac{\partial^2 L_W(\theta\mathbf{1},\mathbf{y})}{\partial\theta^2}$ is not defined at $\theta = \mu(\mathbf{y})$ because $\frac{\partial L_W(\theta\mathbf{1},\mathbf{y})}{\partial\theta}$ is not defined at this point.

The function $L_W(\theta\mathbf{1},\mathbf{y})$ has two global minima at

$$\hat{\theta} = \mu(\mathbf{y}) \pm \sigma(\mathbf{y}) \quad \text{(D.46)}$$

because

(i) The function $L_W(\theta\mathbf{1},\mathbf{y})$ is a continuous function of $\theta$ and

$$\left.\frac{\partial L_W(\theta\mathbf{1},\mathbf{y})}{\partial\theta}\right|_{\theta=\hat{\theta}} = 0 \quad \text{(D.47)}$$

(ii) It has two local minima at $\hat{\theta}$, because

$$L_W(\hat{\theta}\mathbf{1},\mathbf{y}) = (\sigma(\mathbf{y}))/(\sigma(\mathbf{y}) + \text{mad}(\mathbf{y})) < 1 \quad \text{(D.48)}$$

$$\left.\frac{\partial^2 L_W(\theta\mathbf{1},\mathbf{y})}{\partial\theta^2}\right|_{\theta=\hat{\theta}} = \text{mad}(\mathbf{y})/(\sigma(\mathbf{y})(\sigma(\mathbf{y}) + \text{mad}(\mathbf{y}))^2) > 0 \quad \text{(D.49)}$$

(iii) Its value at $\theta = \mu(\mathbf{y})$ equals

$$L_W(\mu(\mathbf{y})\mathbf{1},\mathbf{y}) = 1 \quad \text{(D.50)}$$

(iv) The limits equal

$$\lim_{\theta\to\infty} L_W(\theta\mathbf{1},\mathbf{y}) = \lim_{\theta\to-\infty} L_W(\theta\mathbf{1},\mathbf{y}) = 1 \quad \text{(D.51)}$$

(v) Therefore, the two local minima of $L_W(\theta\mathbf{1},\mathbf{y})$ are global minima for $\theta > \mu(\mathbf{y})$ and $\theta < \mu(\mathbf{y})$ respectively. Since the value of the function $L_W(\theta\mathbf{1},\mathbf{y})$ at the two local minima is equal, then those local minima should be also global minima in the set of real numbers. ■

**Proof D.4** (Consistency of estimators based on $L_W$): We prove that $\hat{\theta}_+(\underline{\mathbf{y}}) = \mu(\underline{\mathbf{y}}) + \sigma(\underline{\mathbf{y}})$ is a consistent estimator of $\mathbb{E}_F[\underline{y}] + \mathbb{V}_F^{1/2}[\underline{y}]$. First, we note that $\mu(\underline{\mathbf{y}})$ converges in probability to $\mathbb{E}_F[\underline{y}]$ for IID $\underline{y}_1, \dots, \underline{y}_n$ with CDF $F$ ([46], page 55). Secondly, we note that $\sigma^2(\underline{\mathbf{y}})$ converges in probability to $\mathbb{V}_F[\underline{y}]$ for IID $\underline{y}_1, \dots, \underline{y}_n$ ([46], page 55). By the continuous mapping theorem, $\sigma(\underline{\mathbf{y}})$ converges in probability to $\mathbb{V}_F^{1/2}[\underline{y}]$. By Slutsky's theorem we have that $\hat{\theta}_+(\underline{y}) \xrightarrow{P_\theta} \mathbb{E}_F[\underline{y}] + \mathbb{V}_F^{1/2}[\underline{y}]$, therefore $\hat{\theta}_+(\underline{y})$ is a consistent estimator. The proof for $\hat{\theta}_-(\underline{y})$ is analogous.■

**Proof D.5** (Translation invariance of $L_{\mathrm{NR}_2}$): Substituting translation vectors into (2.13):

$$L_{\mathrm{NR}_2}(\boldsymbol{z} + c\mathbf{1}, \boldsymbol{y} + c\mathbf{1}) = \|(\boldsymbol{z} + c\mathbf{1}) - (\boldsymbol{y} + c\mathbf{1})\|_2^2 / (\|\boldsymbol{z} + c\mathbf{1} - \mathbf{1}\mu(\boldsymbol{y} + c\mathbf{1})\|_2 + \|\mathbf{1}\mu(\boldsymbol{y} + c\mathbf{1}) - \boldsymbol{y} - c\mathbf{1}\|_2)^2 \Rightarrow \quad \text{(D.52)}$$

$$L_{\mathrm{NR}_2}(\boldsymbol{z} + c\mathbf{1}, \boldsymbol{y} + c\mathbf{1}) = \|\boldsymbol{z} - \boldsymbol{y}\|_2^2 / (\|\boldsymbol{z} - \mathbf{1}\mu(\boldsymbol{y})\|_2 + \|\mathbf{1}\mu(\boldsymbol{y}) - \boldsymbol{y}\|_2)^2 \Rightarrow \quad \text{(D.53)}$$

$$L_{\mathrm{NR}_2}(\boldsymbol{z} + c\mathbf{1}, \boldsymbol{y} + c\mathbf{1}) = L_{\mathrm{NR}_2}(\boldsymbol{z}, \boldsymbol{y}), c \in \mathbb{R} \quad \text{(D.54)}$$

Thus $L_{\mathrm{NR}_2}$ is translation invariant.■

**Proof D.6** ($L_{\mathrm{NR}_2}$ for extremum estimation): We show that $\operatorname*{argmin}_{\theta \in \Theta} L_{\mathrm{NR}_2}(\theta\mathbf{1}, \boldsymbol{y}) = \mu(\boldsymbol{y}) \pm \sigma(\boldsymbol{y})$. For $\boldsymbol{z} = \theta\mathbf{1}$ and $\boldsymbol{y}$ not constant, the nominator of $L_{\mathrm{NR}_2}$ is (see eq. (D.10))

$$\|\boldsymbol{z} - \boldsymbol{y}\|_2^2 = n(\theta - \mu(\boldsymbol{y}))^2 + n\sigma^2(\boldsymbol{y}) \quad \text{(D.55)}$$

The first part of the denominator is:

$$\|\boldsymbol{z} - \mathbf{1}\mu(\boldsymbol{y})\|_2^2 = \|\theta\mathbf{1} - \mathbf{1}\mu(\boldsymbol{y})\|_2^2 \Rightarrow \quad \text{(D.56)}$$

$$\|\boldsymbol{z} - \mathbf{1}\mu(\boldsymbol{y})\|_2^2 = n(\theta - \mu(\boldsymbol{y}))^2 \quad \text{(D.57)}$$

Therefore $L_{\mathrm{NR}_2}(\theta\mathbf{1}, \boldsymbol{y})$ equals

$$L_{\mathrm{NR}_2}(\theta\mathbf{1}, \boldsymbol{y}) = (n(\theta - \mu(\boldsymbol{y}))^2 + n\sigma^2(\boldsymbol{y})) / (\sqrt{n}|\theta - \mu(\boldsymbol{y})| + \sqrt{n}\sigma(\boldsymbol{y}))^2 \Rightarrow \quad \text{(D.58)}$$

$$L_{\mathrm{NR}_2}(\theta\mathbf{1}, \boldsymbol{y}) = ((\theta - \mu(\boldsymbol{y}))^2 + \sigma^2(\boldsymbol{y})) / (|\theta - \mu(\boldsymbol{y})| + \sigma(\boldsymbol{y}))^2 \quad \text{(D.59)}$$

The first partial derivative of $L_{a_2}(\theta\mathbf{1}, \boldsymbol{y})$ is

$$\frac{\partial L_{\mathrm{NR}_2}(\theta\mathbf{1},\boldsymbol{y})}{\partial\theta} = \begin{cases} 2\sigma(\boldsymbol{y})(\theta - \mu(\boldsymbol{y}) - \sigma(\boldsymbol{y}))/(\sigma(\boldsymbol{y}) + \theta - \mu(\boldsymbol{y}))^3, \theta > \mu(\boldsymbol{y}) \\ 2\sigma(\boldsymbol{y})(\theta - \mu(\boldsymbol{y}) + \sigma(\boldsymbol{y}))/(\sigma(\boldsymbol{y}) + \mu(\boldsymbol{y}) - \theta)^3, \theta < \mu(\boldsymbol{y}) \end{cases} \quad \text{(D.60)}$$

or equivalently

$$\frac{\partial L_{\mathrm{NR}_2}(\theta\mathbf{1},\boldsymbol{y})}{\partial\theta} = 2\sigma(\boldsymbol{y})(\theta - \mu(\boldsymbol{y}) - \operatorname{sign}(\theta - \mu(\boldsymbol{y}))\sigma(\boldsymbol{y}))/(\sigma(\boldsymbol{y}) + |\theta - \mu(\boldsymbol{y})|)^3 \quad \text{(D.61)}$$

The second partial derivative of $L_{a_2}(\theta\mathbf{1}, \boldsymbol{y})$ is

$$\frac{\partial^2 L_{\mathrm{NR}_2}(\theta\mathbf{1},\boldsymbol{y})}{\partial\theta^2} = \begin{cases} (-4\sigma(\boldsymbol{y})(\theta - \mu(\boldsymbol{y}) - 2\sigma(\boldsymbol{y})))/(\sigma(\boldsymbol{y}) + \theta - \mu(\boldsymbol{y}))^4, \theta > \mu(\boldsymbol{y}) \\ (-4\sigma(\boldsymbol{y})(\mu(\boldsymbol{y}) - \theta - 2\sigma(\boldsymbol{y})))/(\sigma(\boldsymbol{y}) + \mu(\boldsymbol{y}) - \theta)^4, \theta < \mu(\boldsymbol{y}) \end{cases} \quad \text{(D.62)}$$

or equivalently

$$\frac{\partial^2 L_{\mathrm{NR}_2}(\theta\mathbf{1},\boldsymbol{y})}{\partial\theta^2} = (4\sigma(\boldsymbol{y})(2\sigma(\boldsymbol{y}) - |\theta - \mu(\boldsymbol{y})|))/(\sigma(\boldsymbol{y}) + |\theta - \mu(\boldsymbol{y})|)^4, \theta \in \mathbb{R} \setminus \{\mu(\boldsymbol{y})\} \quad \text{(D.63)}$$

The derivative $\frac{\partial L_{\mathrm{NR}_2}(\theta\mathbf{1},\boldsymbol{y})}{\partial\theta}$ is not defined at $\theta=\mu(\boldsymbol{y})$, because (see Figure 2)

$$\lim_{\theta\to\mu(\boldsymbol{y})^-}\frac{\partial L_{\mathrm{NR}_2}(\theta\mathbf{1},\boldsymbol{y})}{\partial\theta}=2/\sigma(\boldsymbol{y}) \tag{D.64}$$

and

$$\lim_{\theta\to\mu(\boldsymbol{y})^+}\frac{\partial L_{\mathrm{NR}_2}(\theta\mathbf{1},\boldsymbol{y})}{\partial\theta}=-2/\sigma(\boldsymbol{y}) \tag{D.65}$$

The function $L_{\mathrm{NR}_2}(\theta\mathbf{1},\boldsymbol{y})$ has two global minima at

$$\hat{\theta}=\mu(\boldsymbol{y})\pm\sigma(\boldsymbol{y}) \tag{D.66}$$

because

(i) The function $L_{\mathrm{NR}_2}(\theta\mathbf{1},\boldsymbol{y})$ is a continuous function of $\theta$ and

$$\left.\frac{\partial L_{\mathrm{NR}_2}(\theta\mathbf{1},\boldsymbol{y})}{\partial\theta}\right|_{\theta=\hat{\theta}}=0 \tag{D.67}$$

(ii) It has two local minima at $\hat{\theta}$, because

$$L_{\mathrm{NR}_2}(\hat{\theta}\mathbf{1},\boldsymbol{y})=1/2<1 \tag{D.68}$$

$$\left.\frac{\partial^2 L_{\mathrm{NR}_2}(\theta\mathbf{1},\boldsymbol{y})}{\partial\theta^2}\right|_{\theta=\hat{\theta}}=1/(4\sigma^2(\boldsymbol{y}))>0 \tag{D.69}$$

(iii) Its value at $\theta=\mu(\boldsymbol{y})$ equals

$$L_{\mathrm{NR}_2}(\mu(\boldsymbol{y})\mathbf{1},\boldsymbol{y})=1 \tag{D.70}$$

(iv) The limits equal

$$\lim_{\theta\to\infty}L_{\mathrm{NR}_2}(\theta\mathbf{1},\boldsymbol{y})=\lim_{\theta\to-\infty}L_{\mathrm{NR}_2}(\theta\mathbf{1},\boldsymbol{y})=1 \tag{D.71}$$

(v) Therefore, the two local minima of $L_{\mathrm{NR}_2}(\theta\mathbf{1},\boldsymbol{y})$ are global minima for $\theta>\mu(\boldsymbol{y})$ and $\theta<\mu(\boldsymbol{y})$ respectively. Since the values of the function $L_{\mathrm{NR}_2}(\theta\mathbf{1},\boldsymbol{y})$ at the two local minima are equal, then those local minima should be also global minima in the set of real numbers.■

**Proof D.7** ($L_{\mathrm{NR}_2}$ for extremum estimation of a linear model): We show that $\underset{\boldsymbol{\theta}\in\mathbb{R}^2}{\operatorname{argmin}}\,L_{\mathrm{NR}_2}(a\boldsymbol{x}+b\mathbf{1},\boldsymbol{y})=(\operatorname{sign}(\rho(\boldsymbol{x},\boldsymbol{y}))(\|\boldsymbol{y}_c\|_2/\|\boldsymbol{x}_c\|_2),\mu(\boldsymbol{y})-(\operatorname{sign}(\rho(\boldsymbol{x},\boldsymbol{y}))\mu(\boldsymbol{x})\|\boldsymbol{y}_c\|_2)/\|\boldsymbol{x}_c\|_2)$, where $\boldsymbol{x}_c$ and $\boldsymbol{y}_c$ are centered vectors defined in eq. (A.23) and $\rho(\boldsymbol{x},\boldsymbol{y})$ is Pearson correlation defined in eq. (A.28).

We assume that $\boldsymbol{y}$ is not constant and $\boldsymbol{z}=a\boldsymbol{x}+b\mathbf{1}$. Then due to the translation invariance property (eq. (3.12)):

$$L_{\mathrm{NR}_2}(\boldsymbol{z},\boldsymbol{y})=L_{\mathrm{NR}_2}(\boldsymbol{z}-\mu(\boldsymbol{y})\mathbf{1},\boldsymbol{y}-\mu(\boldsymbol{y})\mathbf{1})\Rightarrow \tag{D.72}$$

$$L_{\mathrm{NR}_2}(\boldsymbol{z},\boldsymbol{y}) = L_{\mathrm{NR}_2}(a\boldsymbol{x} + b\mathbf{1} - \mu(\boldsymbol{y})\mathbf{1}, \boldsymbol{y}_c) \tag{D.73}$$

We set:

$$c = a\mu(\boldsymbol{x}) + b - \mu(\boldsymbol{y}) \tag{D.74}$$

Therefore, the first argument of the left part function in eq. (D.73), after substituting b with its transformation in eq. (D.74) is

$$a\boldsymbol{x} + b\mathbf{1} - \mu(\boldsymbol{y})\mathbf{1} = a\boldsymbol{x} + (c - a\mu(\boldsymbol{x}) + \mu(\boldsymbol{y}))\mathbf{1} - \mu(\boldsymbol{y})\mathbf{1} \Rightarrow \tag{D.75}$$

$$a\boldsymbol{x} + b\mathbf{1} - \mu(\boldsymbol{y})\mathbf{1} = a\boldsymbol{x} + c\mathbf{1} - a\mu(\boldsymbol{x})\mathbf{1} \Rightarrow \tag{D.76}$$

$$a\boldsymbol{x} + b\mathbf{1} - \mu(\boldsymbol{y})\mathbf{1} = a\boldsymbol{x}_c + c\mathbf{1} \tag{D.77}$$

Therefore, the arguments of $L_{\mathrm{NR}_2}$ are transformed to include centered vectors that ease the analytical procedure to find the minima:

$$L_{\mathrm{NR}_2}(\boldsymbol{z},\boldsymbol{y}) = L_{\mathrm{NR}_2}(a\boldsymbol{x}_c + c\mathbf{1}, \boldsymbol{y}_c) \Rightarrow \tag{D.78}$$

$$L_{\mathrm{NR}_2}(\boldsymbol{z},\boldsymbol{y}) = \|a\boldsymbol{x}_c + c\mathbf{1} - \boldsymbol{y}_c\|_2^2 / (\|a\boldsymbol{x}_c + c\mathbf{1}\|_2 + \|\boldsymbol{y}_c\|_2)^2 \tag{D.79}$$

The $L_{\mathrm{NR}_2}$ nominator equals

$$\|a\boldsymbol{x}_c + c\mathbf{1} - \boldsymbol{y}_c\|_2^2 = \langle (a\boldsymbol{x}_c - \boldsymbol{y}_c) + c\mathbf{1}, (a\boldsymbol{x}_c - \boldsymbol{y}_c) + c\mathbf{1} \rangle \Rightarrow \tag{D.80}$$

$$\|a\boldsymbol{x}_c + c\mathbf{1} - \boldsymbol{y}_c\|_2^2 = \|a\boldsymbol{x}_c - \boldsymbol{y}_c\|_2^2 + nc^2 + 2\langle a\boldsymbol{x}_c - \boldsymbol{y}_c, c\mathbf{1} \rangle \tag{D.81}$$

By leveraging orthogonality to unity by the eq. (A.27), applied to the last term in the right part of eq. (D.81) we have

$$\|a\boldsymbol{x}_c + c\mathbf{1} - \boldsymbol{y}_c\|_2^2 = \|a\boldsymbol{x}_c - \boldsymbol{y}_c\|_2^2 + nc^2 \Rightarrow \tag{D.82}$$

$$\|a\boldsymbol{x}_c + c\mathbf{1} - \boldsymbol{y}_c\|_2^2 = a^2\|\boldsymbol{x}_c\|_2^2 + \|\boldsymbol{y}_c\|_2^2 - 2a\langle \boldsymbol{x}_c, \boldsymbol{y}_c \rangle + nc^2 \tag{D.83}$$

Furthermore, the first part of the denominator of $L_{\mathrm{NR}_2}$ in eq. (D.79) equals

$$\|a\boldsymbol{x}_c + c\mathbf{1}\|_2^2 = \langle a\boldsymbol{x}_c + c\mathbf{1}, a\boldsymbol{x}_c + c\mathbf{1} \rangle \Rightarrow \tag{D.84}$$

$$\|a\boldsymbol{x}_c + c\mathbf{1}\|_2^2 = a^2\|\boldsymbol{x}_c\|_2^2 + nc^2 \tag{D.85}$$

We set $w_1$ equal to the left part of eq. (D.83) and $w_2$ equal to the left part of eq. (D.85)

$$w_1 = a^2\|\boldsymbol{x}_c\|_2^2 + \|\boldsymbol{y}_c\|_2^2 - 2a\langle \boldsymbol{x}_c, \boldsymbol{y}_c \rangle + nc^2 \tag{D.86}$$

$$w_2 = a^2\|\boldsymbol{x}_c\|_2^2 + nc^2 \tag{D.87}$$

We also set $w_3$ equal to

$$w_3 = w_2 + \|\boldsymbol{y}_c\|_2^2 + 2\|\boldsymbol{y}_c\|_2\sqrt{w_2} \tag{D.88}$$

Therefore,

$$w_1 = w_2 + \|\boldsymbol{y}_c\|_2^2 - 2a\langle \boldsymbol{x}_c, \boldsymbol{y}_c \rangle \tag{D.89}$$

and,

$$L_{\mathrm{NR}_2}(\boldsymbol{z}, \boldsymbol{y}) = w_1/w_3 \tag{D.90}$$

or equivalently

$$L_{\mathrm{NR}_2}(\boldsymbol{z}, \boldsymbol{y}) = (w_2 + \|\boldsymbol{y}_c\|_2^2 - 2a\langle \boldsymbol{x}_c, \boldsymbol{y}_c\rangle)/(w_2 + \|\boldsymbol{y}_c\|_2^2 + 2\|\boldsymbol{y}_c\|_2\sqrt{w_2}) \tag{D.91}$$

Partial derivatives:

The first order partial derivative of $L_{a_2}$ with respect to $c$ is

$$\frac{\partial L_{\mathrm{NR}_2}(\boldsymbol{z},\boldsymbol{y})}{\partial c} = (2ncw_3 - w_1(2nc + 2nc\|\boldsymbol{y}_c\|_2/\sqrt{w_2}))/w_3^2 \Rightarrow \tag{D.92}$$

$$\frac{\partial L_{\mathrm{NR}_2}(\boldsymbol{z},\boldsymbol{y})}{\partial c} = 2nc(\sqrt{w_2}w_3 - \sqrt{w_2}w_1 - w_1\|\boldsymbol{y}_c\|_2)/(\sqrt{w_2}w_3^2) \tag{D.93}$$

or equivalently

$$\frac{\partial L_{\mathrm{NR}_2}(\boldsymbol{z},\boldsymbol{y})}{\partial c} = (2ncw_4)/(\sqrt{w_2}w_3^2) \tag{D.94}$$

where

$$w_4 = 2a\langle \boldsymbol{x}_c, \boldsymbol{y}_c\rangle(\sqrt{a^2\|\boldsymbol{x}_c\|_2^2 + nc^2} + \|\boldsymbol{y}_c\|_2) + \|\boldsymbol{y}_c\|_2(a^2\|\boldsymbol{x}_c\|_2^2 + nc^2 - \|\boldsymbol{y}_c\|_2^2) \tag{D.95}$$

or equivalently

$$w_4 = 2a\langle \boldsymbol{x}_c, \boldsymbol{y}_c\rangle(\sqrt{w_2} + \|\boldsymbol{y}_c\|_2) + \|\boldsymbol{y}_c\|_2(w_2 - \|\boldsymbol{y}_c\|_2^2) \Rightarrow \tag{D.96}$$

$$w_4 = (\sqrt{w_2} + \|\boldsymbol{y}_c\|_2)(2a\langle \boldsymbol{x}_c, \boldsymbol{y}_c\rangle + \|\boldsymbol{y}_c\|_2(\sqrt{w_2} - \|\boldsymbol{y}_c\|_2)) \tag{D.97}$$

The first order partial derivative of $L_{\mathrm{NR}_2}$ with respect to $a$ is

$$\frac{\partial L_{\mathrm{NR}_2}(\boldsymbol{z},\boldsymbol{y})}{\partial a} = (2(a\|\boldsymbol{x}_c\|_2^2 - \langle \boldsymbol{x}_c, \boldsymbol{y}_c\rangle)w_3) - (2a\|\boldsymbol{x}_c\|_2^2 w_1(1 + \|\boldsymbol{y}_c\|_2/\sqrt{w_2}))/w_3^2 \Rightarrow \tag{D.98}$$

$$\frac{\partial L_{\mathrm{NR}_2}(\boldsymbol{z},\boldsymbol{y})}{\partial a} = w_5/(\sqrt{w_2}w_3^2) \tag{D.99}$$

where

$$w_5 = 2\sqrt{w_2}(a\|\boldsymbol{x}_c\|_2^2 - \langle \boldsymbol{x}_c, \boldsymbol{y}_c\rangle)(\sqrt{w_2} + \|\boldsymbol{y}_c\|_2)^2 - 2a\|\boldsymbol{x}_c\|_2^2(w_2 + \|\boldsymbol{y}_c\|_2^2 - 2a\langle \boldsymbol{x}_c, \boldsymbol{y}_c\rangle)(\sqrt{w_2} + \|\boldsymbol{y}_c\|_2) \Rightarrow \tag{D.100}$$

$$w_5 = (\sqrt{w_2} + \|\boldsymbol{y}_c\|_2)(2\sqrt{w_2}(a\|\boldsymbol{x}_c\|_2^2 - \langle \boldsymbol{x}_c, \boldsymbol{y}_c\rangle)(\sqrt{w_2} + \|\boldsymbol{y}_c\|_2) - 2a\|\boldsymbol{x}_c\|_2^2(w_2 + \|\boldsymbol{y}_c\|_2^2 - 2a\langle \boldsymbol{x}_c, \boldsymbol{y}_c\rangle)) \tag{D.101}$$

Minimization:

To minimize $L_{\mathrm{NR}_2}(a\boldsymbol{x} + b\boldsymbol{1}, \boldsymbol{y})$ we observe from eq. (D.94) that the partial derivative

$\frac{\partial L_{\mathrm{NR}_2}(\boldsymbol{z},\boldsymbol{y})}{\partial c}$ equals 0 if and only if $c = 0$ or $w_4 = 0$.

<u>Case 1 ($c = 0$):</u>

$$c = 0 \Leftrightarrow a\mu(\boldsymbol{x}) + b - \mu(\boldsymbol{y}) = 0 \Leftrightarrow b = \mu(\boldsymbol{y}) - a\mu(\boldsymbol{x}) \quad \text{(D.102)}$$

We have to find solutions of the equation $\frac{\partial L_{\mathrm{NR}_2}(\boldsymbol{z},\boldsymbol{y})}{\partial a} = 0$ (where the partial derivative is defined in eq. (D.99)), i.e. equivalently solutions of the equation $w_5 = 0$. First, we compute $w_2$ from eq. (D.87):

$$w_2 = a^2\|\boldsymbol{x}_c\|_2^2 \quad \text{(D.103)}$$

Then we compute $w_5$ from eq. (D.101):

$$w_5 = (|a|\|\boldsymbol{x}_c\|_2 + \|\boldsymbol{y}_c\|_2)(2|a|\|\boldsymbol{x}_c\|_2(a\|\boldsymbol{x}_c\|_2^2 - \langle\boldsymbol{x}_c,\boldsymbol{y}_c\rangle)(|a|\|\boldsymbol{x}_c\|_2 + \|\boldsymbol{y}_c\|_2) - 2a\|\boldsymbol{x}_c\|_2^2(a^2\|\boldsymbol{x}_c\|_2^2 + \|\boldsymbol{y}_c\|_2^2 - 2a\langle\boldsymbol{x}_c,\boldsymbol{y}_c\rangle)) \Rightarrow \quad \text{(D.104)}$$

$$w_5 = 2\|\boldsymbol{x}_c\|_2(|a|\|\boldsymbol{x}_c\|_2 + \|\boldsymbol{y}_c\|_2)((a\|\boldsymbol{x}_c\|_2^2 - \langle\boldsymbol{x}_c,\boldsymbol{y}_c\rangle)(a^2\|\boldsymbol{x}_c\|_2 + |a|\|\boldsymbol{y}_c\|_2) - \|\boldsymbol{x}_c\|_2(a^3\|\boldsymbol{x}_c\|_2^2 + a\|\boldsymbol{y}_c\|_2^2 - 2a^2\langle\boldsymbol{x}_c,\boldsymbol{y}_c\rangle)) \Rightarrow \quad \text{(D.105)}$$

$$w_5 = 2\|\boldsymbol{x}_c\|_2(|a|\|\boldsymbol{x}_c\|_2 + \|\boldsymbol{y}_c\|_2)((a^3\|\boldsymbol{x}_c\|_2^3 + a|a|\|\boldsymbol{x}_c\|_2^2\|\boldsymbol{y}_c\|_2 - a^2\|\boldsymbol{x}_c\|_2\langle\boldsymbol{x}_c,\boldsymbol{y}_c\rangle - |a|\|\boldsymbol{y}_c\|_2\langle\boldsymbol{x}_c,\boldsymbol{y}_c\rangle) - a^3\|\boldsymbol{x}_c\|_2^3 - a\|\boldsymbol{x}_c\|_2\|\boldsymbol{y}_c\|_2^2 + 2a^2\|\boldsymbol{x}_c\|_2\langle\boldsymbol{x}_c,\boldsymbol{y}_c\rangle) \Rightarrow \quad \text{(D.106)}$$

$$w_5 = 2\|\boldsymbol{x}_c\|_2(|a|\|\boldsymbol{x}_c\|_2 + \|\boldsymbol{y}_c\|_2)(a|a|\|\boldsymbol{x}_c\|_2^2\|\boldsymbol{y}_c\|_2 - |a|\|\boldsymbol{y}_c\|_2\langle\boldsymbol{x}_c,\boldsymbol{y}_c\rangle - a\|\boldsymbol{x}_c\|_2\|\boldsymbol{y}_c\|_2^2 + a^2\|\boldsymbol{x}_c\|_2\langle\boldsymbol{x}_c,\boldsymbol{y}_c\rangle) \quad \text{(D.107)}$$

$w_5$ is a function of $|a|$, so to find where it equals 0 we examine two cases for $a$. The first case is $a > 0$, when $w_5$ simplifies to:

$$w_5 = 2a\|\boldsymbol{x}_c\|_2(a\|\boldsymbol{x}_c\|_2 + \|\boldsymbol{y}_c\|_2)(a\|\boldsymbol{x}_c\|_2^2\|\boldsymbol{y}_c\|_2 - \|\boldsymbol{y}_c\|_2\langle\boldsymbol{x}_c,\boldsymbol{y}_c\rangle - \|\boldsymbol{x}_c\|_2\|\boldsymbol{y}_c\|_2^2 + a\|\boldsymbol{x}_c\|_2\langle\boldsymbol{x}_c,\boldsymbol{y}_c\rangle) \quad \text{(D.108)}$$

Therefore, $w_5 = 0$ when:

$$a = (\|\boldsymbol{y}_c\|_2\langle\boldsymbol{x}_c,\boldsymbol{y}_c\rangle + \|\boldsymbol{x}_c\|_2\|\boldsymbol{y}_c\|_2^2)/(\|\boldsymbol{x}_c\|_2^2\|\boldsymbol{y}_c\|_2 + \|\boldsymbol{x}_c\|_2\langle\boldsymbol{x}_c,\boldsymbol{y}_c\rangle) \Rightarrow \quad \text{(D.109)}$$

$$a = \|\boldsymbol{y}_c\|_2/\|\boldsymbol{x}_c\|_2 \quad \text{(D.110)}$$

The first case is $a < 0$, when $w_5$ simplifies to:

$$w_5 = 2a\|\boldsymbol{x}_c\|_2(a\|\boldsymbol{x}_c\|_2 + \|\boldsymbol{y}_c\|_2)(-a\|\boldsymbol{x}_c\|_2^2\|\boldsymbol{y}_c\|_2 + \|\boldsymbol{y}_c\|_2\langle\boldsymbol{x}_c,\boldsymbol{y}_c\rangle - \|\boldsymbol{x}_c\|_2\|\boldsymbol{y}_c\|_2^2 + a\|\boldsymbol{x}_c\|_2\langle\boldsymbol{x}_c,\boldsymbol{y}_c\rangle) \quad \text{(D.111)}$$

Therefore, $w_5 = 0$ when:

$$a = (-\|\boldsymbol{y}_c\|_2\langle\boldsymbol{x}_c,\boldsymbol{y}_c\rangle + \|\boldsymbol{x}_c\|_2\|\boldsymbol{y}_c\|_2^2)/(-\|\boldsymbol{x}_c\|_2^2\|\boldsymbol{y}_c\|_2 + \|\boldsymbol{x}_c\|_2\langle\boldsymbol{x}_c,\boldsymbol{y}_c\rangle) \Rightarrow \quad \text{(D.112)}$$

$$a = -\|\boldsymbol{y}_c\|_2/\|\boldsymbol{x}_c\|_2 \quad \text{(D.113)}$$

Summarizing the two cases ($a > 0$ and $a < 0$), we have that $w_5 = 0$ for values of $a = \hat{a}$, where $\hat{a}$:

$$\hat{a} = \pm\|\boldsymbol{y}_c\|_2/\|\boldsymbol{x}_c\|_2 \tag{D.114}$$

From eq. (D.102), after substituting $a$ with $\hat{a}$, we obtain the respective values of $b$:

$$\hat{b} = \mu(\boldsymbol{y}) \mp \mu(\boldsymbol{x})\|\boldsymbol{y}_c\|_2/\|\boldsymbol{x}_c\|_2 \tag{D.115}$$

The value of $L_{\mathrm{NR}_2}(\boldsymbol{z},\boldsymbol{y})$ at $(\hat{a},\hat{b})$ can be computed from eq. (D.90) after first computing $w_2$ (based on eqs. (D.87) and (D.114)) and $w_3$ (based on eqs. (D.88) and (D.114)):

$$w_2 = \|\boldsymbol{y}_c\|_2^2 \tag{D.116}$$

$$w_3 = 4\|\boldsymbol{y}_c\|_2^2 \tag{D.117}$$

Therefore $L_{\mathrm{NR}_2}(\boldsymbol{z},\boldsymbol{y})$ at $(\hat{a},\hat{b})$ equals

$$L_{\mathrm{NR}_2}(\boldsymbol{z},\boldsymbol{y}) = (\|\boldsymbol{y}_c\|_2^2 - \hat{a}\langle\boldsymbol{x}_c,\boldsymbol{y}_c\rangle)/(2\|\boldsymbol{y}_c\|_2^2) \Rightarrow \tag{D.118}$$

$$L_{\mathrm{NR}_2}(\boldsymbol{z},\boldsymbol{y}) = (\|\boldsymbol{y}_c\|_2^2 \mp (\|\boldsymbol{y}_c\|_2/\|\boldsymbol{x}_c\|_2)\langle\boldsymbol{x}_c,\boldsymbol{y}_c\rangle)/(2\|\boldsymbol{y}_c\|_2^2) \Rightarrow \tag{D.119}$$

$$L_{\mathrm{NR}_2}(\boldsymbol{z},\boldsymbol{y}) = (1 \mp \rho(\boldsymbol{x},\boldsymbol{y}))/2 \tag{D.120}$$

Therefore $L_{\mathrm{NR}_2}$ has two local optima. Which one of them is the minimum depends on the sign of $\rho(\boldsymbol{x},\boldsymbol{y})$. If $\rho(\boldsymbol{x},\boldsymbol{y}) > 0$ then $L_{\mathrm{NR}_2}(\boldsymbol{z},\boldsymbol{y})$ attains its minimum at $(1 - \rho(\boldsymbol{x},\boldsymbol{y})/2$ when

$$\hat{a} = \|\boldsymbol{y}_c\|_2/\|\boldsymbol{x}_c\|_2 \tag{D.121}$$

If $\rho(\boldsymbol{x},\boldsymbol{y}) < 0$ then $L_{\mathrm{NR}_2}(\boldsymbol{z},\boldsymbol{y})$ attains its minimum at $(1 + \rho(\boldsymbol{x},\boldsymbol{y})/2$ when

$$\hat{a} = -\|\boldsymbol{y}_c\|_2/\|\boldsymbol{x}_c\|_2 \tag{D.122}$$

Those two cases can be combined to obtain a unique estimate for $a$:

$$\hat{a} = \mathrm{sign}(\rho(\boldsymbol{x},\boldsymbol{y}))(\|\boldsymbol{y}_c\|_2/\|\boldsymbol{x}_c\|_2) \tag{D.123}$$

From eq. (D.102), after substituting $a$ with $\hat{a}$, we obtain the respective values of $b$:

$$\hat{b} = \mu(\boldsymbol{y}) - (\mathrm{sign}(\rho(\boldsymbol{x},\boldsymbol{y}))\mu(\boldsymbol{x})\|\boldsymbol{y}_c\|_2)/\|\boldsymbol{x}_c\|_2 \tag{D.124}$$

At $(\hat{a},\hat{b})$, $L_{\mathrm{NR}_2}(\boldsymbol{z},\boldsymbol{y})$ attains its minimum value

$$L_{\mathrm{NR}_2}(\boldsymbol{z},\boldsymbol{y}) = (1 - |\rho(\boldsymbol{x},\boldsymbol{y})|)/2 \tag{D.125}$$

To complete the proof, one possible way is to compute the second derivatives. However this is complicated. Yet, the global optimality of $L_{\mathrm{NR}_2}(\boldsymbol{z},\boldsymbol{y})$ at $c = 0$ can also be proven by observing that for fixed $a$:

(i) $L_{\mathrm{NR}_2}(a\boldsymbol{x}_c + c\boldsymbol{1},\boldsymbol{y}_c)$ is even in $c$, as it depends on $c$ only through $w_2$ (see eqs. (D.90), (D.89) and (D.88)):

$$L_{\mathrm{NR}_2}(a\boldsymbol{x}_c + c\mathbf{1}, \boldsymbol{y}_c) = L_{\mathrm{NR}_2}(a\boldsymbol{x}_c - c\mathbf{1}, \boldsymbol{y}_c) \quad \text{(D.126)}$$

because from eq. (D.87), $w_2 = a^2\|\boldsymbol{x}_c\|_2^2 + nc^2$.

(ii) It is straightforward to prove that:

$$\lim_{c\to\infty} L_{\mathrm{NR}_2}(a\boldsymbol{x}_c + c\mathbf{1}, \boldsymbol{y}_c) = \lim_{c\to-\infty} L_{\mathrm{NR}_2}(a\boldsymbol{x}_c + c\mathbf{1}, \boldsymbol{y}_c) = 1 \geq (1 - |\rho(\boldsymbol{x}, \boldsymbol{y})|)/2 \quad \text{(D.127)}$$

(iii) $L_{\mathrm{NR}_2}(a\boldsymbol{x}_c + c\mathbf{1}, \boldsymbol{y}_c)$ is continuous and differentiable for every $c$.

(iv) $c = 0$ is a unique critical point since $\frac{\partial L_{\mathrm{NR}_2}(\boldsymbol{z},\boldsymbol{y})}{\partial c} \neq 0$, otherwise (see case 2 that proves that $w_4 \neq 0$).

Due to properties (iii) and (iv), there is no other local optimum, therefore $c = 0$ should be a global optimum. Then property (ii) implies that $c = 0$ is a global minimum.

To prove the global optimality for $\hat{a}$, we note that from eqs. (D.87) – (D.90), at $c = 0$:

$$L_{\mathrm{NR}_2}(\boldsymbol{z}, \boldsymbol{y}) = \|a\boldsymbol{x}_c - \boldsymbol{y}_c\|_2^2/(\|a\boldsymbol{x}_c\|_2 + \|\boldsymbol{y}_c\|_2)^2 \Rightarrow \quad \text{(D.128)}$$

$$L_{\mathrm{NR}_2}(\boldsymbol{z}, \boldsymbol{y}) = (a^2\|\boldsymbol{x}_c\|_2^2 + \|\boldsymbol{y}_c\|_2^2 - 2a\langle\boldsymbol{x}_c, \boldsymbol{y}_c\rangle)/(a^2\|\boldsymbol{x}_c\|_2^2 + \|\boldsymbol{y}_c\|_2^2 + 2|a|\|\boldsymbol{x}_c\|_2\|\boldsymbol{y}_c\|_2) \Rightarrow \quad \text{(D.129)}$$

$$L_{\mathrm{NR}_2}(\boldsymbol{z}, \boldsymbol{y}) = (a^2\|\boldsymbol{x}_c\|_2^2 + \|\boldsymbol{y}_c\|_2^2 - 2a\rho(\boldsymbol{x}, \boldsymbol{y})\|\boldsymbol{x}_c\|_2\|\boldsymbol{y}_c\|_2)/(a^2\|\boldsymbol{x}_c\|_2^2 + \|\boldsymbol{y}_c\|_2^2 + 2|a|\|\boldsymbol{x}_c\|_2\|\boldsymbol{y}_c\|_2) \Rightarrow \quad \text{(D.130)}$$

$$L_{\mathrm{NR}_2}(\boldsymbol{z}, \boldsymbol{y}) = ((a/\hat{a})^2 + 1 - 2(a/\hat{a})|\rho(\boldsymbol{x}, \boldsymbol{y})|)/((a/\hat{a})^2 + 1 + 2\mathrm{sign}(\rho(\boldsymbol{x}, \boldsymbol{y}))|a|/\hat{a}) \quad \text{(D.131)}$$

We set

$$t = a/\hat{a} \quad \text{(D.132)}$$

Then we have four cases:

(i) If $a > 0$ and $\rho(\boldsymbol{x}, \boldsymbol{y}) > 0$, then $t > 0$ and:

$$L_{\mathrm{NR}_2}(\boldsymbol{z}, \boldsymbol{y}) = 1 - (2t(1 + \rho(\boldsymbol{x}, \boldsymbol{y})))/(t + 1)^2 \quad \text{(D.133)}$$

This is minimized for $t = 1$, i.e. when $a = \hat{a}$ and $L_{\mathrm{NR}_2}(\boldsymbol{z}, \boldsymbol{y}) = (1 - |\rho(\boldsymbol{x}, \boldsymbol{y})|)/2$, which is a global minimum.

(ii) If $a < 0$ and $\rho(\boldsymbol{x}, \boldsymbol{y}) > 0$, then $t < 0$ and:

$$L_{\mathrm{NR}_2}(\boldsymbol{z}, \boldsymbol{y}) = 1 + (2t(1 - \rho(\boldsymbol{x}, \boldsymbol{y})))/(t - 1)^2 \quad \text{(D.134)}$$

This is minimized for $t = -1$, i.e. when $a = \hat{a}$ and $L_{\mathrm{NR}_2}(\boldsymbol{z}, \boldsymbol{y}) = (1 + |\rho(\boldsymbol{x}, \boldsymbol{y})|)/2$, which is not a global minimum.

(iii) If $a > 0$ and $\rho(\boldsymbol{x}, \boldsymbol{y}) < 0$, then $t < 0$ and:

$$L_{\mathrm{NR}_2}(\boldsymbol{z}, \boldsymbol{y}) = 1 + (2t(1 + \rho(\boldsymbol{x}, \boldsymbol{y})))/(t-1)^2 \tag{D.135}$$

This is minimized for $t = -1$, i.e. when $a = \hat{a}$ and $L_{\mathrm{NR}_2}(\boldsymbol{z}, \boldsymbol{y}) = (1 + |\rho(\boldsymbol{x}, \boldsymbol{y})|)/2$, which is not a global minimum.

(iv) If $a < 0$ and $\rho(\boldsymbol{x}, \boldsymbol{y}) < 0$, then $t > 0$ and:

$$L_{\mathrm{NR}_2}(\boldsymbol{z}, \boldsymbol{y}) = 1 - (2t(1 - \rho(\boldsymbol{x}, \boldsymbol{y})))/(t+1)^2 \tag{D.136}$$

This is minimized for $t = 1$, i.e. when $a = \hat{a}$ and $L_{\mathrm{NR}_2}(\boldsymbol{z}, \boldsymbol{y}) = (1 - |\rho(\boldsymbol{x}, \boldsymbol{y})|)/2$, which is a global minimum.

This completes the proof for the case of $a$.

<u>Case 2 ($w_4 = 0$):</u>

We prove that $L_{\mathrm{NR}_2}(\boldsymbol{z}, \boldsymbol{y}) > (1 - |\rho(\boldsymbol{x}, \boldsymbol{y})|)/2$, when $w_4 = 0$, where $w_4$ is defined in eq. (D.97). We try to solve the equation:

$$w_4 = 0 \Rightarrow \tag{D.137}$$

$$(\sqrt{w_2} + \|\boldsymbol{y}_c\|_2)\big(2a\langle \boldsymbol{x}_c, \boldsymbol{y}_c\rangle + \|\boldsymbol{y}_c\|_2(\sqrt{w_2} - \|\boldsymbol{y}_c\|_2)\big) = 0 \Rightarrow \tag{D.138}$$

$$-2a\langle \boldsymbol{x}_c, \boldsymbol{y}_c\rangle = \|\boldsymbol{y}_c\|_2(\sqrt{w_2} - \|\boldsymbol{y}_c\|_2) \tag{D.139}$$

This can also be written as:

$$\sqrt{w_2} = \|\boldsymbol{y}_c\|_2 - 2a\rho(\boldsymbol{x}, \boldsymbol{y})\|\boldsymbol{x}_c\|_2 \tag{D.140}$$

Therefore, from eq. (D.89):

$$w_1 = w_2 + \|\boldsymbol{y}_c\|_2^2 + \|\boldsymbol{y}_c\|_2(\sqrt{w_2} - \|\boldsymbol{y}_c\|_2) \Rightarrow \tag{D.141}$$

$$w_1 = w_2 + \|\boldsymbol{y}_c\|_2\sqrt{w_2} \Rightarrow \tag{D.142}$$

$$w_1 = \sqrt{w_2}(\sqrt{w_2} + \|\boldsymbol{y}_c\|_2) \tag{D.143}$$

Furthermore, from eq. (D.88):

$$w_3 = (\sqrt{w_2} + \|\boldsymbol{y}_c\|_2)^2 \tag{D.144}$$

Returning to eq. (D.90) and substituting the values of $w_1$, $\sqrt{w_2}$ and $w_3$:

$$L_{\mathrm{NR}_2}(\boldsymbol{z}, \boldsymbol{y}) = \sqrt{w_2}/(\sqrt{w_2} + \|\boldsymbol{y}_c\|_2) \Rightarrow \tag{D.145}$$

$$L_{\mathrm{NR}_2}(\boldsymbol{z}, \boldsymbol{y}) = (\|\boldsymbol{y}_c\|_2 - 2a\rho(\boldsymbol{x}, \boldsymbol{y})\|\boldsymbol{x}_c\|_2)/(2(\|\boldsymbol{y}_c\|_2 - a\rho(\boldsymbol{x}, \boldsymbol{y})\|\boldsymbol{x}_c\|_2)) \tag{D.146}$$

We set,

$$w_6 = (a\|\boldsymbol{x}_c\|_2)/\|\boldsymbol{y}_c\|_2 \tag{D.147}$$

and substitute in eq. (D.146):

$$L_{\mathrm{NR}_2}(\boldsymbol{z},\boldsymbol{y}) = (1-2\rho(\boldsymbol{x},\boldsymbol{y})w_6)/(2(1-\rho(\boldsymbol{x},\boldsymbol{y})w_6)) \tag{D.148}$$

Furthermore, the following inequalities hold (based on eqs. (D.87), (D.140) and (D.147)):

$$\sqrt{w_2} \geq |a|\|\boldsymbol{x}_{\mathrm{c}}\|_2 \geq 0 \Rightarrow \tag{D.149}$$

$$\|\boldsymbol{y}_c\|_2 - 2a\rho(\boldsymbol{x},\boldsymbol{y})\|\boldsymbol{x}_{\mathrm{c}}\|_2 \geq |a|\|\boldsymbol{x}_{\mathrm{c}}\|_2 \geq 0 \Rightarrow \tag{D.150}$$

$$1-2\rho(\boldsymbol{x},\boldsymbol{y})w_6 \geq |w_6| \geq 0 \tag{D.151}$$

Now we have to prove that $L_{\mathrm{NR}_2}(\boldsymbol{z},\boldsymbol{y}) > (1-|\rho(\boldsymbol{x},\boldsymbol{y})|)/2$ or equivalently (based on eq. (D.148)) that:

$$(1-2\rho(\boldsymbol{x},\boldsymbol{y})w_6)/(2(1-\rho(\boldsymbol{x},\boldsymbol{y})w_6)) \geq (1-|\rho(\boldsymbol{x},\boldsymbol{y})|)/2 \tag{D.152}$$

or equivalently

$$(1-2\rho(\boldsymbol{x},\boldsymbol{y})w_6)/(1-\rho(\boldsymbol{x},\boldsymbol{y})w_6) \geq 1-|\rho(\boldsymbol{x},\boldsymbol{y})| \Rightarrow \tag{D.153}$$

$$(-\rho(\boldsymbol{x},\boldsymbol{y})w_6 + |\rho(\boldsymbol{x},\boldsymbol{y})| - \rho(\boldsymbol{x},\boldsymbol{y})|\rho(\boldsymbol{x},\boldsymbol{y})|w_6)/(1-\rho(\boldsymbol{x},\boldsymbol{y})w_6) \geq 0 \tag{D.154}$$

We examine two cases for proving the inequality in eq. (D.154), depending on the sign of $\rho(\boldsymbol{x},\boldsymbol{y})$:

<u>Case 2.a:</u>

If $\rho(\boldsymbol{x},\boldsymbol{y}) > 0$ we have to prove:

$$(\rho(\boldsymbol{x},\boldsymbol{y})(-w_6 + 1 - \rho(\boldsymbol{x},\boldsymbol{y})w_6))/(1-\rho(\boldsymbol{x},\boldsymbol{y})w_6) \geq 0 \Rightarrow \tag{D.155}$$

$$(\rho(\boldsymbol{x},\boldsymbol{y})(1 - w_6(1+\rho(\boldsymbol{x},\boldsymbol{y}))))/(1-\rho(\boldsymbol{x},\boldsymbol{y})w_6) \geq 0 \tag{D.156}$$

(i) Sign of the denominator:

If $1-\rho(\boldsymbol{x},\boldsymbol{y})w_6 \leq 0$, then $w_6 \geq 1/\rho(\boldsymbol{x},\boldsymbol{y})$. However, the constraint $1-2\rho(\boldsymbol{x},\boldsymbol{y})w_6 \geq |w_6|$ would imply $1 \geq (2\rho(\boldsymbol{x},\boldsymbol{y})+1)w_6$, which in turn would imply that $1 \geq (2\rho(\boldsymbol{x},\boldsymbol{y})+1)/\rho(\boldsymbol{x},\boldsymbol{y})$ or equivalently $-1 \geq 1/\rho(\boldsymbol{x},\boldsymbol{y})$ which in turn is impossible, therefore $1-\rho(\boldsymbol{x},\boldsymbol{y})w_6 > 0$.

(ii) Sign of the nominator:

To prove that the nominator is positive, it suffices to prove that $w_6 \leq 1/(1+\rho(\boldsymbol{x},\boldsymbol{y}))$.

If $w_6 \geq 0$, then $1-2\rho(\boldsymbol{x},\boldsymbol{y})w_6 \geq |w_6|$ implies that $w_6 \leq 1/(1+2\rho(\boldsymbol{x},\boldsymbol{y}))$, which implies that $w_6 \leq 1/(1+\rho(\boldsymbol{x},\boldsymbol{y}))$.

If $w_6 < 0$, then $w_6 < 0 \leq 1/(1+\rho(\boldsymbol{x},\boldsymbol{y}))$.

<u>Case 2.b:</u>

If $\rho(\boldsymbol{x},\boldsymbol{y}) < 0$ we have to prove:

$$(-\rho(\boldsymbol{x},\boldsymbol{y})(w_6 + 1 - \rho(\boldsymbol{x},\boldsymbol{y})w_6))/(1-\rho(\boldsymbol{x},\boldsymbol{y})w_6) \geq 0 \Rightarrow \tag{D.157}$$

$$(|\rho(\boldsymbol{x},\boldsymbol{y})|(1+(1+|\rho(\boldsymbol{x},\boldsymbol{y})|)w_6))/(1+|\rho(\boldsymbol{x},\boldsymbol{y})|w_6) \geq 0 \quad (D.158)$$

(i) Sign of the denominator:

If $w_6 \geq 0$, then $1+|\rho(\boldsymbol{x},\boldsymbol{y})|w_6 \geq 0$.

If $w_6 < 0$, then $1-2\rho(\boldsymbol{x},\boldsymbol{y})w_6 \geq |w_6|$ implies that $w_6 \geq -1/(2|\rho(\boldsymbol{x},\boldsymbol{y})|+1)$, which in turn implies that $w_6 \geq -1/(2|\rho(\boldsymbol{x},\boldsymbol{y})|+1) \geq -1/|\rho(\boldsymbol{x},\boldsymbol{y})|$, therefore $1+|\rho(\boldsymbol{x},\boldsymbol{y})|w_6 \geq 0$.

(ii) Sign of the nominator:

To prove that the nominator is positive, it suffices to prove that $w_6 \geq -1/(1+|\rho(\boldsymbol{x},\boldsymbol{y})|)$.

If $w_6 > 0$, then $w_6 > 0 > -1/(1+|\rho(\boldsymbol{x},\boldsymbol{y})|)$.

If $w_6 < 0$, then we have proved that $w_6 \geq -1/(2|\rho(\boldsymbol{x},\boldsymbol{y})|+1)$, that implies $w_6 \geq -1/(2|\rho(\boldsymbol{x},\boldsymbol{y})|+1) \geq -1/(1+|\rho(\boldsymbol{x},\boldsymbol{y})|)$ that completes the proof.■

## Appendix E Statistical software

All analyses were conducted using the `R` programming language [31].

Data processing and visualization were performed using the `data.table` [47] and `tidyverse` ([48], [49]) packages.

Hydrologic modeling was carried out with the `airGR` package ([33], [34]), which we modified to implement the SE, $L_{\mathrm{NR}_2}$ and $L_W$ loss functions as additional calibration options. The required hydrologic data were retrieved from the `airGRdatasets` package [35].

Model performance metrics were computed using the `scoringfunctions` package ([50], [51]).

Finally, all reports were generated using the `devtools` [52], `knitr` ([53], [54], [55]) and `rmarkdown` ([56], [57], [58]) packages.

**Supplementary material**: We provide the following files to reproduce results of Sections 3 and 4:

Section 3.1.3: `01_figure_Lw_extremum_estimator.Rmd`, `01_figure_Lw_extremum_estimator.html`.

Section 3.2.4: `02_figure_LNR2_extremum_estimator.Rmd`, `02_figure_LNR2_extremum_estimator.html`.

Section 4.1: `03_figure_forecasting_Gaussian_distribution.Rmd`,

`03_figure_forecasting_Gaussian_distribution.html`.

Section 4.2: `04a_linear_model.Rmd`, `04a_linear_model.html`, `04b_linear_model.Rmd`, `04b_linear_model.html`, `04c_linear_model.Rmd`, `04c_linear_model.html`.

Section 4.3: `05a_hydro_data.Rmd`, `05a_hydro_data.html`, `05a_hydro_data.RData`, `05b_hydro_modelling.Rmd`, `05b_hydro_modelling.html`, `05b_hydro_modelling.RData`, `05c_hydro_modelling.Rmd`, `05c_hydro_modelling.html`, `05d_hydro_modelling.Rmd`, `05d_hydro_modelling.html`.

**Conflicts of interest:** The authors declare no conflict of interest.

**Statement:** During the preparation of this work, the authors used DeepSeek-V.3.2 to assist with language polishing and to enhance readability. After using this service, the authors reviewed and edited the content as needed and take full responsibility for the content of the published article.

## References

[1] Gneiting T (2011) Making and evaluating point forecasts. Journal of the American Statistical Association 106(494):746–762. https://doi.org/10.1198/jasa.2011.r10138.

[2] Bennett ND, Croke BFW, Guariso G, Guillaume JHA, Hamilton SH, Jakeman AJ, Marsili-Libelli S, Newham LTH, Norton JP, Perrin C, Pierce SA, Robson B, Seppelt R, Voinov AA, Fath BD, Andreassian V (2013) Characterising performance of environmental models. Environmental Modelling and Software 40:1–20. https://doi.org/10.1016/j.envsoft.2012.09.011.

[3] Jackson EK, Roberts W, Nelsen B, Williams GP, Nelson EJ, Ames DP (2019) Introductory overview: Error metrics for hydrologic modelling – A review of common practices and an open source library to facilitate use and adoption. Environmental Modelling and Software 119:32–48. https://doi.org/10.1016/j.envsoft.2019.05.001.

[4] Moriasi DN, Arnold JG, Van Liew MW, Bingner RL, Harmel RD, Veith TL (2007) Model evaluation guidelines for systematic quantification of accuracy in watershed simulations. Transactions of the ASABE 50(3):885–900. https://doi.org/10.13031/2013.23153.

[5] Willmott CJ (1981) On the validation of models. Physical Geography 2(2):184–194. https://doi.org/10.1080/02723646.1981.10642213.

[6] Willmott CJ (1982) Some comments on the evaluation of model performance. Bulletin of the American Meteorological Society 63(11):1309–1313. https://doi.org/10.1175/1520-0477(1982)063<1309:SCOTEO>2.0.CO;2.

[7] Newey WK, McFadden D (1994) Large sample estimation and hypothesis testing. In: Engle RF, McFadden D (eds) Handbook of Econometrics, vol. 4. Elsevier, Amsterdam, pp 2111–2245. https://doi.org/10.1016/S1573-4412(05)80005-4.
[8] Legates DR, McCabe Jr GJ (1999) Evaluating the use of "goodness-of-fit" Measures in hydrologic and hydroclimatic model validation. Water Resources Research 35(1):233–241. https://doi.org/10.1029/1998WR900018.
[9] Krause P, Boyle DP, Bäse F (2005) Comparison of different efficiency criteria for hydrological model assessment. Advances in Geosciences 5:89–97. https://doi.org/10.5194/adgeo-5-89-2005.
[10] Dimitriadis T, Fissler T, Ziegel J (2024) Characterizing *M*-estimators. Biometrika 111(1):339–346. https://doi.org/10.1093/biomet/asad026.
[11] Murphy AH, Daan H (1985) Forecast evaluation. In: Murphy AH, Katz RW (eds) Probability, Statistics and Decision Making in the Atmospheric Sciences. CRC Press, pp 379–437.
[12] Banerjee A, Guo X, Wang H (2005) On the optimality of conditional expectation as a Bregman predictor. IEEE Transactions on Information Theory 51(7):2664–2669. https://doi.org/10.1109/TIT.2005.850145.
[13] Patton AJ (2011) Volatility forecast comparison using imperfect volatility proxies. Journal of Econometrics 160(1):246–256. https://doi.org/10.1016/j.jeconom.2010.03.034.
[14] Reichelstein S, Osband K (1984) Incentives in government contracts. Journal of Econometrics 24(2):257–270. https://doi.org/10.1016/0047-2727(84)90029-X.
[15] Saerens M (2000) Building cost functions minimizing to some summary statistics. IEEE Transactions on Neural Networks 11(6):1263–1271. https://doi.org/10.1109/72.883416.
[16] Savage LJ (1971) Elicitation of personal probabilities and expectations. Journal of the American Statistical Association 66(336):783–810. https://doi.org/10.1080/01621459.1971.10482346.
[17] Nash JE, Sutcliffe JV (1970) River flow forecasting through conceptual models part I — A discussion of principles. Journal of Hydrology 10(3):282–290. https://doi.org/10.1016/0022-1694(70)90255-6.
[18] Gneiting T, Resin J (2023) Regression diagnostics meets forecast evaluation: Conditional calibration, reliability diagrams, and coefficient of determination. Electronic Journal of Statistics 17(2):3226–3286. https://doi.org/10.1214/23-EJS2180.
[19] Nolde N, Ziegel JF (2017) Elicitability and backtesting: Perspectives for banking regulation. Annals of Applied Statistics 11(4):1833–1874. https://doi.org/10.1214/17-AOAS1041.
[20] Fissler T, Lorentzen C, Mayer M (2023) Model comparison and calibration assessment: User guide for consistent scoring functions in machine learning and actuarial practice. https://arxiv.org/abs/2202.12780.
[21] Fissler T, Ziegel JF (2016) Higher order elicitability and Osband's principle. The Annals of Statistics 44(4):1680–1707. https://doi.org/10.1214/16-AOS1439.
[22] Osband KH (1985) Providing Incentives for Better Cost Forecasting. PhD thesis, University of California, Berkeley. https://doi.org/10.5281/zenodo.4355667.
[23] Steinwart I, Pasin C, Williamson R, Zhang S (2014) Elicitation and identification of properties. Proceedings of Machine Learning Research 35:482–526.

[24] Patton AJ (2020) Comparing possibly misspecified forecasts. Journal of Business and Economic Statistics 38(4):796–809. https://doi.org/10.1080/07350015.2019.1585256.
[25] Amemiya T (1973) Regression analysis when the dependent variable is truncated normal. Econometrica 41(6):997–1016. https://doi.org/10.2307/1914031.
[26] Amemiya T (1985) Advanced Econometrics. Cambridge, MA: Harvard University Press. ISBN: 9780674251991.
[27] Huber PJ (1964) Robust estimation of a location parameter. The Annals of Mathematical Statistics 35(1):73–101. https://doi.org/10.1214/aoms/1177703732.
[28] Huber PJ (1967) The behavior of maximum likelihood estimates under nonstandard conditions. In: Le Cam LM, Nyman J (eds) Proceedings of the Fifth Berkeley Symposium on Mathematical Statistics and Probability. Berkeley: University of California Press, Berkeley, pp 221–233.
[29] Willmott CJ, Ackleson SG, Davis RE, Feddema JJ, Klink KM, Legates DR, O'Donnell J, Rowe CM (1985) Statistics for the evaluation of model performance. Journal of Geophysical Research: Oceans 90(C5):8995–9005. https://doi.org/10.1029/JC090iC05p08995.
[30] Chen Z (1996) Conditional $L_p$-quantiles and their application to the testing of symmetry in non-parametric regression. Statistics and Probability Letters 29(2):107–115. https://doi.org/10.1016/0167-7152(95)00163-8.
[31] R Core Team (2025) R: A language and environment for statistical computing. R Foundation for Statistical Computing, Vienna, Austria. https://www.R-project.org.
[32] Perrin C, Michel C, Andréassian V (2003) Improvement of a parsimonious model for streamflow simulation. Journal of Hydrology 279(1–4):275–289. https://doi.org/10.1016/S0022-1694(03)00225-7.
[33] Coron L, Thirel G, Delaigue O, Perrin C, Andréassian V (2017) The suite of lumped GR hydrological models in an R package. Environmental Modelling and Software 94:166–171. https://doi.org/10.1016/j.envsoft.2017.05.002.
[34] Coron L, Delaigue O, Thirel G, Dorchies D, Perrin C, Michel C (2023) airGR: Suite of GR Hydrological Models for Precipitation-Runoff Modelling. R package version 1.7.6. https://doi.org/10.32614/CRAN.package.airGR.
[35] Delaigue O, Brigode P, Thirel G (2023) airGRdatasets: Hydro-Meteorological Catchments Datasets for the 'airGR' Packages. R package version 0.2.1. https://doi.org/10.32614/CRAN.package.airGRdatasets.
[36] Koenker RW, Bassett Jr G (1978) Regression quantiles. Econometrica 46(1):33–50. https://doi.org/10.2307/1913643.
[37] Pande S (2013a) Quantile hydrologic model selection and model structure deficiency assessment: 1. Theory. Water Resources Research 49(9):5631–5657. https://doi.org/10.1002/wrcr.20411.
[38] Pande S (2013b) Quantile hydrologic model selection and model structure deficiency assessment: 2. Applications. Water Resources Research 49(9):5658–5673. https://doi.org/10.1002/wrcr.20422.
[39] Tyralis H, Papacharalampous G (2021) Quantile-based hydrological modelling. Water 13(23):3420. https://doi.org/10.3390/w13233420.
[40] Newey WK, Powell JL (1987) Asymmetric least squares estimation and testing. Econometrica 55(4):819–847. https://doi.org/10.2307/1911031.

[41] Tyralis H, Papacharalampous G, Khatami S (2023) Expectile-based hydrological modelling for uncertainty estimation: Life after mean. Journal of Hydrology 617(Part B):128986. https://doi.org/10.1016/j.jhydrol.2022.128986.
[42] Osband K, Reichelstein S (1985) Information-eliciting compensation schemes. Journal of Public Economics 27(1):107–115. https://doi.org/10.1016/0047-2727(85)90031-3.
[43] Vrugt J (2024) Distribution-based model evaluation and diagnostics: Elicitability, propriety, and scoring rules for hydrograph functionals. Water Resources Research 60(6):e2023WR036710. https://doi.org/10.1029/2023WR036710.
[44] Gentle JE (2024) Matrix Algebra. Springer Cham. https://doi.org/10.1007/978-3-031-42144-0.
[45] Muroi H, Adachi S (2015) Model validation criteria for system identification in time domain. IFAC-PapersOnLine 48(28):86–91. https://doi.org/10.1016/j.ifacol.2015.12.105.
[46] Lehmann EL, Casella G (1998) Theory of Point Estimation. Springer New York, NY. https://doi.org/10.1007/b98854.
[47] Barrett T, Dowle M, Srinivasan A, Gorecki J, Chirico M, Hocking T, Schwendinger B, Krylov I (2025) data.table: Extension of 'data.frame'. R package version 1.17.8. https://doi.org/10.32614/CRAN.package.data.table.
[48] Wickham H, Averick M, Bryan J, Chang W, McGowan LD, François R, Grolemund G, Hayes A, Henry L, Hester J, Kuhn M, Pedersen TL, Miller E, Bache SM, Müller K, Ooms J, Robinson D, Paige Seidel DP, Spinu V, Takahashi K, Vaughan D, Wilke C, Woo K, Yutani H (2019) Welcome to the tidyverse. Journal of Open Source Software 4(43):1686. https://doi.org/10.21105/joss.01686.
[49] Wickham H (2023) tidyverse: Easily install and load the 'Tidyverse'. R package version 2.0.0. https://doi.org/10.32614/CRAN.package.tidyverse.
[50] Tyralis H, Papacharalampous G (2024) A review of predictive uncertainty estimation with machine learning. Artificial Intelligence Review 57(94). https://doi.org/10.1007/s10462-023-10698-8.
[51] Tyralis H, Papacharalampous G (2025) scoringfunctions: A Collection of Loss Functions for Assessing Point Forecasts. R package version 1.1. https://doi.org/10.32614/CRAN.package.scoringfunctions.
[52] Wickham H, Hester J, Chang W, Bryan J (2022) devtools: Tools to make developing R packages easier. R package version 2.4.5. https://doi.org/10.32614/CRAN.package.devtools.
[53] Xie Y (2014) knitr: A Comprehensive Tool for Reproducible Research in R. In: Stodden V, Leisch F, Peng RD (Eds) Implementing Reproducible Computational Research. Chapman and Hall/CRC.
[54] Xie Y (2015) Dynamic Documents with R and knitr, 2nd edition. Chapman and Hall/CRC.
[55] Xie Y (2025) knitr: A general-purpose package for dynamic report generation in R. R package version 1.50. https://doi.org/10.32614/CRAN.package.knitr.
[56] Allaire JJ, Xie Y, Dervieux C, McPherson J, Luraschi J, Ushey K, Atkins A, Wickham H, Cheng J, Chang W, Iannone R (2024) rmarkdown: Dynamic documents for R. R package version 2.29. https://doi.org/10.32614/CRAN.package.rmarkdown.
[57] Xie Y, Allaire JJ, Grolemund G (2018) R Markdown: The Definitive Guide. Chapman and Hall/CRC. ISBN 9781138359338. https://bookdown.org/yihui/rmarkdown.

[58] Xie Y, Dervieux C, Riederer E (2020) R Markdown Cookbook. Chapman and Hall/CRC. ISBN 9780367563837. https://bookdown.org/yihui/rmarkdown-cookbook.